\documentclass[showpacs,amsmath,twocolumn,showkeys,pra]{revtex4}
\usepackage{dcolumn}    
\usepackage{bm}         
\usepackage{ifpdf}
\usepackage{amssymb,lineno,amsfonts}
\usepackage{graphicx}   
\usepackage{bbm}
\usepackage{mathrsfs}
\usepackage{upgreek}
\usepackage{epstopdf}
\usepackage{setspace}
\usepackage{float}
\usepackage{hyperref}
\usepackage[matrix,frame,arrow]{xypic}
\usepackage{float}
\usepackage{amsmath}
\usepackage{natbib}
\usepackage{color} 
\usepackage{subfigure}
\definecolor{med-blue}{RGB}{25,25,112} 
\hypersetup{colorlinks, linkcolor={blue},citecolor={blue}, urlcolor={blue}}
\newcommand{\ket}[1]{\vert{#1}\rangle}
\newcommand{\bra}[1]{\langle{#1}\vert}
\newcommand{\outpr}[2]{\vert{#1}\rangle\langle{#2}\vert}

\newcommand{\proj}[1]{\outpr{#1}{#1}}

\newcommand{\tr}{\mathrm{Tr}}

\begin{document}
	\title{Unambiguous measurement of information scrambling in a\\ hierarchical star-topology system}
	\author{Deepak Khurana}
	\email{deepak.khurana@students.iiserpune.ac.in}
	\author{V. R. Krithika}
	\email{krithika\_vr@students.iiserpune.ac.in}
	\author{T. S. Mahesh}
		\email{mahesh.ts@iiserpune.ac.in}
	\affiliation{ Department of Physics and NMR Research Center,\\
		Indian Institute of Science Education and Research, Pune 411008, India}

	\begin{abstract}
		{We investigate the scrambling of  information in a hierarchical star-topology system using out-of-time-ordered correlation (OTOC) functions.  The system consists of a central qubit directly interacting with a set of satellite qubits, which in turn interact with a second layer of satellite qubits.  This particular topology not only allows  convenient preparation and filtering of multiple quantum coherences between the central qubit and the first layer but also to engineer  scrambling in a controlled manner.   Hence, it provides us with an opportunity to experimentally study  scrambling of information localized in multi-spin correlations via the construction of relevant OTOCs. Since the measurement of OTOC requires a time evolution, the non-scrambling processes such as decoherence and certain experimental errors create an ambiguity. Therefore, the unambiguous quantification of information scrambling requires suppressing contributions from decoherence to the OTOC dynamics. To this end, we propose and experimentally demonstrate a constant time protocol which is able to filter contribution exclusively from information scrambling.}
	\end{abstract}
		
\maketitle
\section{Introduction}
\textit{Scrambling} of initially localized quantum information into many degrees of freedom via the creation of non-local correlations leads to a perceived loss of quantum information in practical time scales. In recent investigations, measurement of information scrambling has been related to many practical aspects such as  diagnosis of quantum chaos \cite{shenker2014black,maldacena2016bound,hosur2016chaos}, entanglement \cite{garttner2018relating}, detection of  many-body localization \cite{wei2018exploring,huang2017out,fan2017out,chen2016universal}, quantum phase transitions \cite{heyl2018detecting,banerjee2017solvable}, and thermalization \cite{wei2018emergent}. The center to all these studies is the experimentally measurable physical   quantity called the out-of-time-ordered correlation (OTOC) functions \cite{shenker2014black,larkin1969quasiclassical,maldacena2016bound}. An OTOC function is four point correlation function where the operators are not ordered in time, and its temporal decay is taken as indication of information scrambling in a many-body quantum system \cite{roberts2015localized}.  Despite significant  theoretical investigations across condensed matter and high-energy research, experimental measurement of OTOC functions is challenging because it involves the reversal of time evolution. Several protocols such as  interferometric \cite{swingle2016measuring,yao2016interferometric}, quantum clock \cite{zhu2016measurement}, and quasi-probabilities \cite{halpern2018quasiprobability} are proposed. On the experimental side, early success with nuclear magnetic resonance (NMR) \cite{li2017measuring,niknam2018sensitivity,wei2018exploring} and ion-traps platforms \cite{landsman2019verified,garttner2017measuring} have been reported. 

In realistic scenarios, decoherence and experimental errors also contribute to the decay of OTOC, and thereby create an ambiguity in the observation of information scrambling \cite{swingle2018resilience,garttner2017measuring}. 
To address this issue, methods based on the use of quantum teleportation \cite{yoshida2019disentangling,landsman2019verified} and OTOC quasi-probabilities \cite{alonso2019out} have been put forward recently for verified measurement of information scrambling.  

Till now, experimental studies have largely focused on the investigation  of scrambling of information localized in uncorrelated degrees of freedom. Recently, the scrambling of information localized in many-body correlations, such as multiple quantum coherences (MQCs) has also been reported \cite{niknam2018sensitivity}. MQCs have been used for practical purposes such as quantum sensing \cite{jones2009magnetic}, detecting entanglement \cite{garttner2018relating}, noise spectroscopy \cite{khurana2016spectral} to name a few which makes it imperative to study the impact of scrambling on these states. 

In this regard, star-topology systems are well suited for this purpose because they allow a controlled and convenient  preparation of various MQCs \cite{jones2009magnetic,shuklaNOON}. Further, if the star-topology system has multiple layers, it provides an opportunity to study scrambling of information localized in multi-spin correlations of inner layers to outer layers. Such hierarchical star-topology systems (HSTS) are important not only from the perspective of studying information scrambling but also they can be treated as  a model for realistic environments to study dynamics of open quantum systems. For example, $^{15}$N impurities around a nitrogen vacancy center in diamond acts the first layer and surrounding $^{13}$C spins constitute the second layer \cite{hanson2008coherent}. 

In this work, using nuclear magnetic resonance (NMR) methods, we study information scrambling in a HSTS consisting of a central qubit surrounded by two layers of satellite qubits. We can initialize the system in a desired MQC between the central qubit and the first layer.  Subsequently, we drive the dynamics from the scrambling to non-scrambling regime by tuning the nonintegrability of evolution propagator. Moreover, we propose  a constant time protocol (CTP) to solely capture the scrambling dynamics while disregarding the decoherence effects. Finally, we experimentally demonstrate  the CTP protocol for the exclusive study of scrambling dynamics of a specific MQC. 

The paper is organized as follows: In the following section, we briefly review the OTOC formalism and introduce CTP. In section III, we describe the experimental system and explain MQC preparation. In section IV, we first investigate the OTOC dynamics corresponding to various MQCs by numerical methods. Further, we describe the experimental study of the scrambling dynamics of a particular MQC using CTP. Finally, we conclude in section V.  

\section{OTOC function and its unambiguous Measurement}
Consider two operators $B(t)$ and $A(0)$, with commutator $C(t)=[A(0),B(t)]$ and let $C(0) = 0$.
OTOC function is then defined as \cite{larkin1969quasiclassical,maldacena2016bound,shenker2014black}
\begin{equation}
O(t) = \langle B^{\dagger}(t) A^\dagger(0) B(t) A(0)\rangle_{\beta},
\label{ft}
\end{equation} 
where $B(t) = U^{\dagger}(t)B(0)U(t)$ is evolved in Heisenberg picture with unitary operator $U(t) = e^{-i{\cal H}t}$ with $\hbar = 1$. Here ${\cal H}$ is the Hamiltonian governing the system dynamics and  $\langle * \rangle_{\beta} = \tr (*\cdot e^{-\beta {\cal H}} )/Z$ is the average over a thermal ensemble prepared with a temperature $1/(k_B\beta)$, with $k_B$ being the Boltzmann constant and $Z = \tr(e^{-\beta {\cal H}})$ being the partition function.
If $A(0)$ and $B(t)$ are unitaries, then the OTOC function can be related to the norm
of the commutator $C(t)$ by 
\begin{equation}
{\cal O}(t) = \mathrm{Re}[O(t)] =  1-\frac{1}{2}\left\langle C^\dagger(t) C(t)\right\rangle_\beta.
\label{reft}
\end{equation}

In general, as ${\cal O}(t)$ evolves under the unitary $U(t)$, it exhibits occasional revivals to unity unless there exists a loss of information. This loss of information is either due to decoherence or due to the leakage of information via scrambling.
In either case, the above commutation norm fails to vanish over time, thus preventing the OTOC revivals. 
However, in practice, both of these effects lead to an effective loss of OTOC revivals in practical timescales.


In the following we assume $A(0) = \rho(0)$, the initial state of the system and $B(t)$ is a unitary operator.  Let us consider following two extreme cases.

(i) A pure initial state ($\rho^2(0) = \rho(0)$) corresponding to zero temperature, i.e., $\beta \rightarrow \infty$.  In this case 
\begin{align}
{\cal O}(t) &= \mathrm{Re}[\langle B^{\dagger}(t) \rho^\dagger(0) B(t) \rho(0)\rangle_{\beta \rightarrow \infty}] \nonumber \\
& = \mathrm{Re}[\mathrm{Tr}\{B^{\dagger}(t) \rho(0) B(t) \rho(0)\}].
\end{align}

(ii) A highly mixed qubit state corresponding to high-temperature NMR conditions, $\rho(0) = \mathbbm{1}/2 +\epsilon \rho_\Delta(0)$, where the traceless part $\rho_\Delta(0)$ is often termed as the deviation density matrix.  Here $\epsilon \propto \beta \simeq 0$ is the purity factor.  Now,
\begin{align}
{\cal O}(t)  &= \mathrm{Re}[\langle B^{\dagger}(t) \rho^\dagger(0) B(t) \rho(0)\rangle_{\beta \rightarrow 0}] \nonumber \\
 & \sim \mathrm{Re}[\mathrm{Tr}\{B^{\dagger}(t) \rho_\Delta(0) B(t) \rho_\Delta(0)\}],
 \label{rhodelta}
\end{align}
up to $\epsilon^2$ factor and a constant background (see Appendix \ref{appA}).

%
%
Moreover, if $B^\dagger(t) = U(t)$ is the evolution propagator, then
\begin{align}
{\cal O}(t) & = \langle \rho(t) \vert \rho(0) \rangle ~~\mbox{or}~~ \langle \rho_\Delta(t) \vert \rho_\Delta(0) \rangle
\end{align}
as is relevant.
Thus in this setting,  ${\cal O}(t)$ can be measured by the overlap between the instantaneous state with the initial state.  

\begin{figure}
	\begin{center}
		\includegraphics[trim = 3.5cm 7cm 7cm 3cm, clip, width=9cm ]{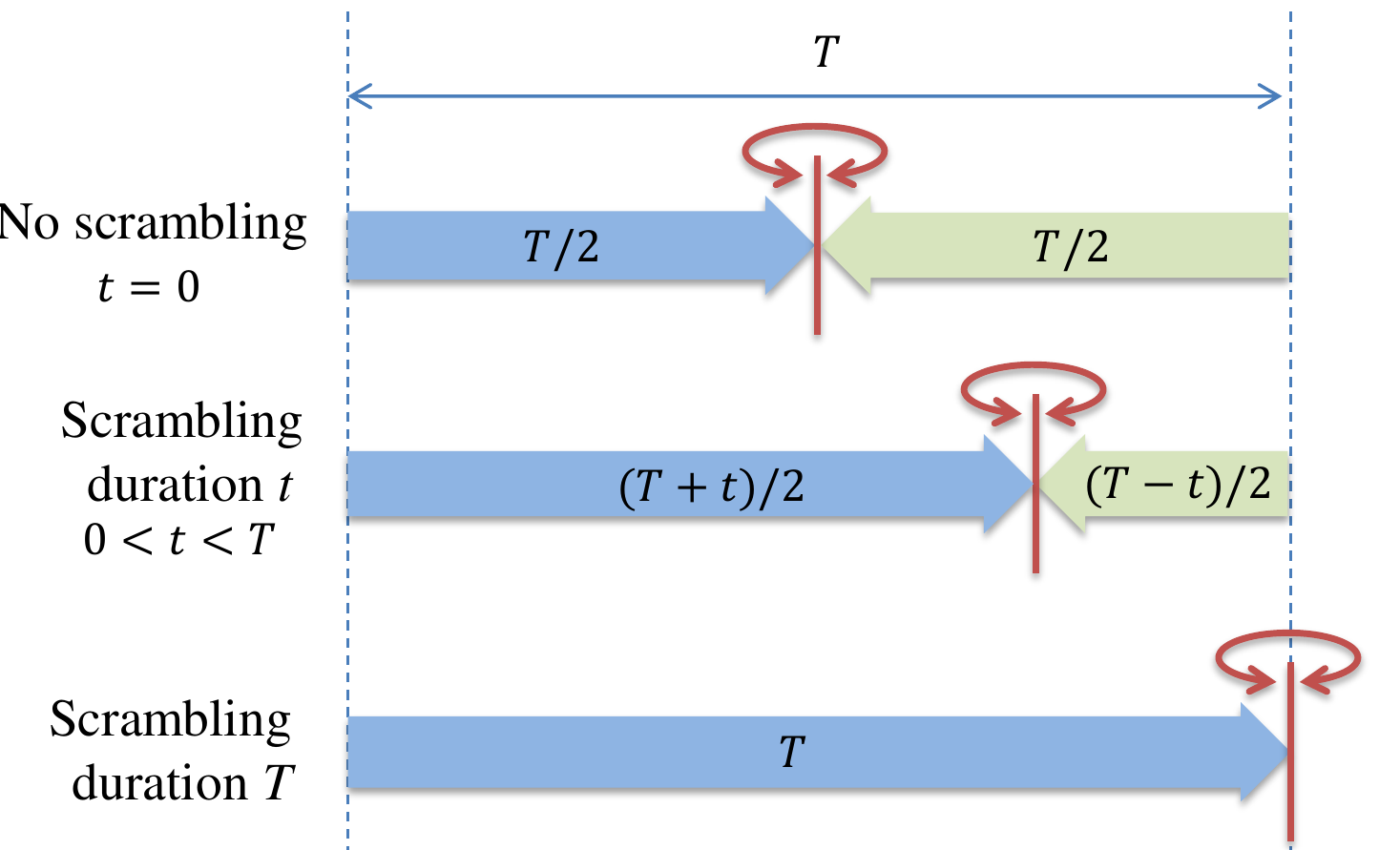}
	\end{center}
	\caption{Schematic illustration of CTP protocol.  The blue and green bars indicate forward and backward evolutions respectively.  While  decoherence is active throughout the duration $T$,  scrambling is active only for the net forward evolution time $t$.} 
	\label{ctp}
\end{figure}

In order to perform an exclusive study of scrambling, it is important to separate the decoherence effects. To this end, certain  protocols based on OTOC quasi probabilities \cite{halpern2018quasiprobability,alonso2019out} and quantum teleportation \cite{landsman2019verified,yoshida2019disentangling} have been proposed.  In the following we propose an alternate approach based on the constant-time protocol (CTP) (illustrated in Fig.\ref{ctp}) commonly used in multi-dimensional NMR spectroscopy \cite{cavanagh1995protein}.

We decompose the time-evolution unitary $U(t)$ into two parts,
\begin{align}
U(t) &= e^{-i{\cal H}t} \nonumber \\ 
& = e^{i{\cal H}(T-t)/2}e^{-i{\cal H}(T+t)/2} \nonumber \\    
& = U^\dagger \left(\frac{T-t}{2}\right)U\left(\frac{T+t}{2}\right).  
\label{cpt}
\end{align}

Thus scrambling under the unitary effectively happens only for time $t$, but the decoherence is active throughout the total time $T$. Hence by carrying out multiple experiments by varying $t$ for an experimentally feasible fixed $T$, one can reconstruct unambiguous evolution under scrambling Hamiltonian. This protocol can be incorporated in all the standard OTOC measurement methods \cite{swingle2016measuring,li2017measuring,garttner2018relating,niknam2018sensitivity,zhu2016measurement,halpern2018quasiprobability,garttner2017measuring,yao2016interferometric}.

\section{Combination MQCs in a HSTS}
\begin{figure}[h]
	\includegraphics[trim = 1cm 3.9cm 1cm 0.8cm, clip, width=7.5cm ]{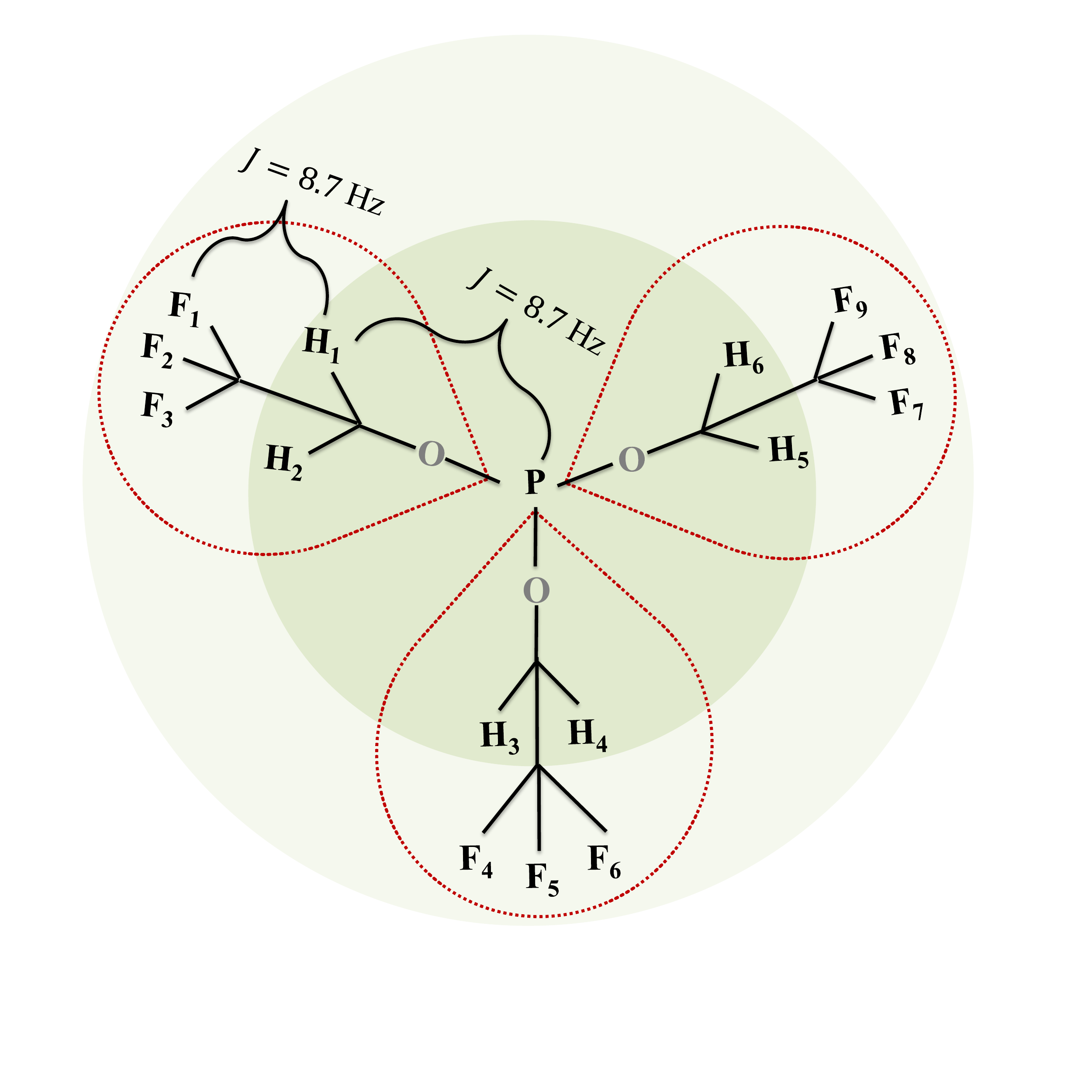}
	\caption{Molecule structure of tris(2,2,2-trifluoroethyl) phosphite. One $^{31}$P spin, six $^{1}$H spins and nine $^{19}$F spins act as central spin, first and second layer respectively. Each branch constitutes  one $^{31}$P spin, two $^{1}$H spins and three $^{19}$F spins.   }
	\label{mol}
\end{figure}

In this work, we consider an  $N$-qubit HSTS  with a central qubit surrounded by $N_1$ qubits in the first layer and $N_2$ qubits in the second layer.
Specifically, the experimental NMR system consists of a $^{31}$P spin  surrounded by a layer of six equivalent  $^{1}$H spins. Each of $^{1}$H spin is further coupled to three $^{19}$F spins in the second layer, as shown in Fig \ref{mol}. Such a system allows us to explore controlled scrambling of information stored in correlations of the central qubit with the first layer to the second layer. 

Let $\hbar = 1$, $\alpha \in \{x,y,z\}$, and $\sigma_\alpha^P$, $\sigma_{i\alpha}^H$ and $\sigma_{j\alpha}^F$ be Pauli-$\alpha$ operators for $^{31}$P, $i$th $^{1}$H and $j$th $^{19}$F respectively.  
We also define the collective terms $$H_\alpha = \sum_{i=1}^{N_1} \sigma^{H}_{i\alpha} ~~\mbox{and}~~
F_\alpha = \sum_{j=1}^{N_2} \sigma^{F}_{j\alpha}.$$
The Hamiltonian of a $K$-branch system can be written as a sum of internal interactions and external fields,  
\begin{equation}
{\cal H}_K = \sum_{k=1}^{K} {\cal H}_{\mathrm{int}}^{(k)} + {\cal H}_{\mathrm{ext}}^{(k)}.
\label{Hamil}
\end{equation}
The branch-wise decomposition of the Hamiltonian is convenient for numerical simulations of OTOC dynamics with partial system size.  
Here the $k$th branch internal interaction Hamiltonian is   
\begin{equation}
{\cal H}_{\mathrm{int}}^{(k)} =  \frac{\pi J}{2}\left(\sum_{i = 1}^{2}\sigma_{z}^{P}\sigma_{mz}^{H} +\sum_{i = 1}^{2}\sum_{j = 1}^{3} \sigma_{mz}^{H} \sigma_{nz}^{F}\right), 
\label{internal}
\end{equation}
where $m = 2(k-1)+i$ and $n = 3(k-1)+j$. Thus, in each branch, the central $^{31}$P spin is coupled to two $^1$H spins and each $^1$H spin is further coupled to three $^{19}$F spins. In our system, $J = 8.7$ Hz happens to be the single scalar coupling constant. 

The external Hamiltonian ${\cal H}_{\mathrm{ext}}^{(k)}$ on the $k$th branch constitutes the application of equal amplitudes $gJ$ of $x$ and $z$ fields employed to introduce non-integrability in the dynamics:
\begin{equation}
{\cal H}_{\mathrm{ext}}^{(k)} = \frac{gJ \pi}{2}\sum_{\alpha\in{x,z}}
\left(\sigma_{\alpha}^P + \sum_{i = 1}^{2}\sigma_{m\alpha}^H + \sum_{j = 1}^{3}\sigma_{n\alpha}^F\right).
\label{xzfield}
\end{equation}



\begin{figure*}
	\subfigure{	\includegraphics[trim = 3cm 9.8cm 3.8cm 9.4cm, clip, width=7.2cm ]{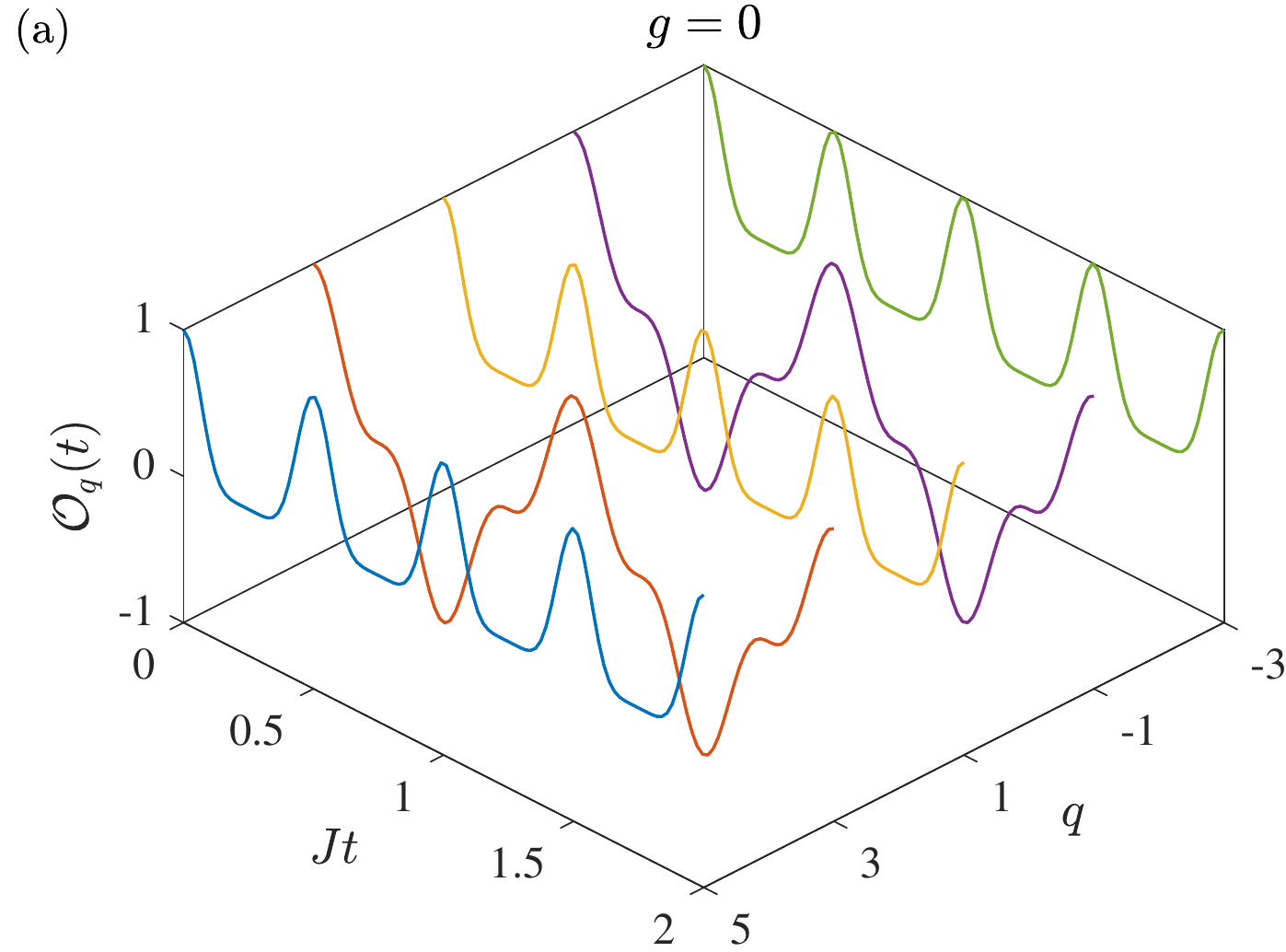}}\hspace{1cm}
		\subfigure{	\includegraphics[trim = 3cm 9.8cm 3.8cm 9.4cm, clip, width=7.2cm ]{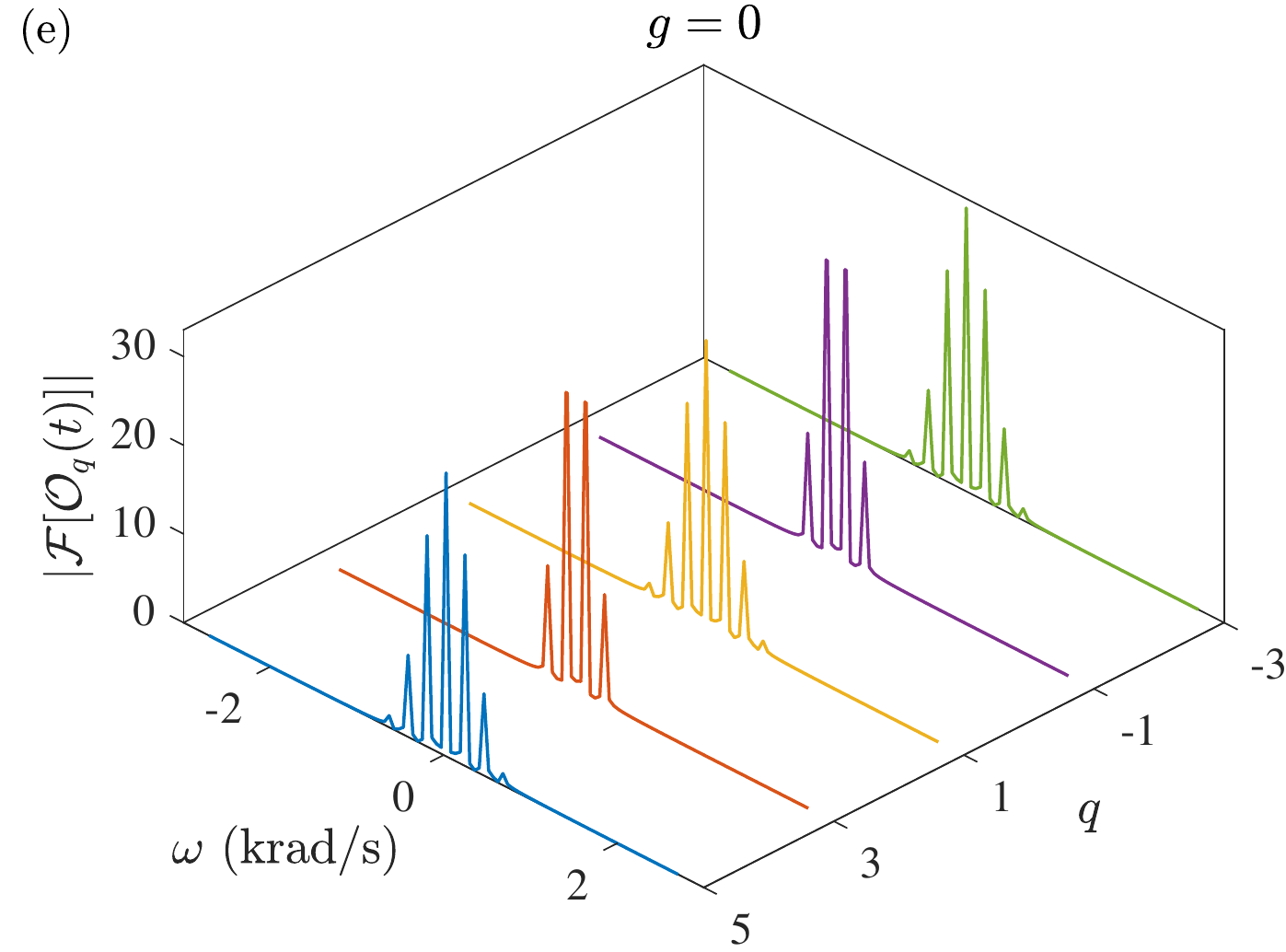}}\hspace{1cm}\\
	\subfigure{
		\includegraphics[trim = 3cm 9.8cm 3.8cm 9.4cm, clip, width=7.2cm ]{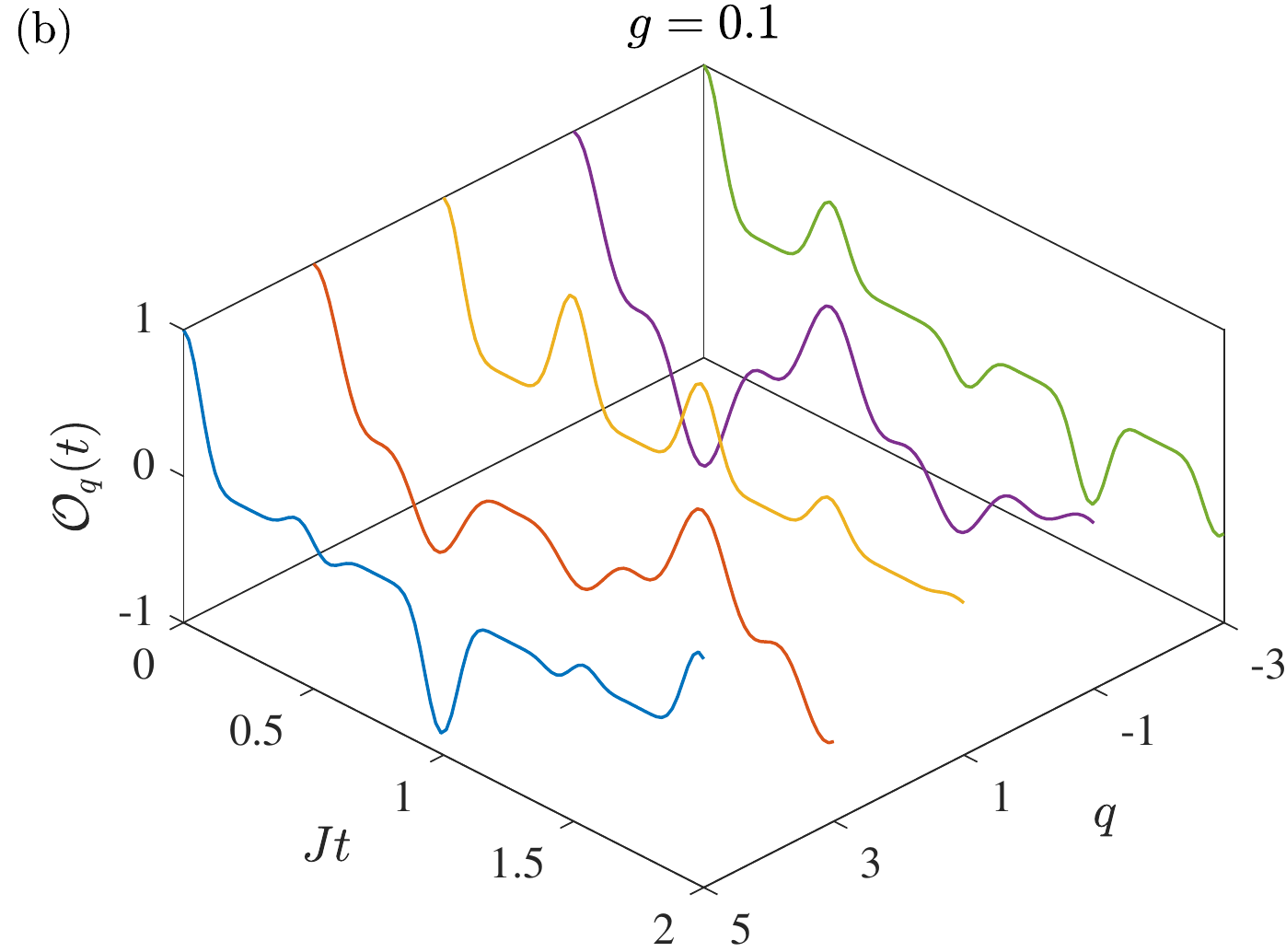}
	}\hspace{0.5cm}
	\subfigure{
	\includegraphics[trim = 3cm 9.8cm 3.8cm 9.4cm, clip, width=7.2cm ]{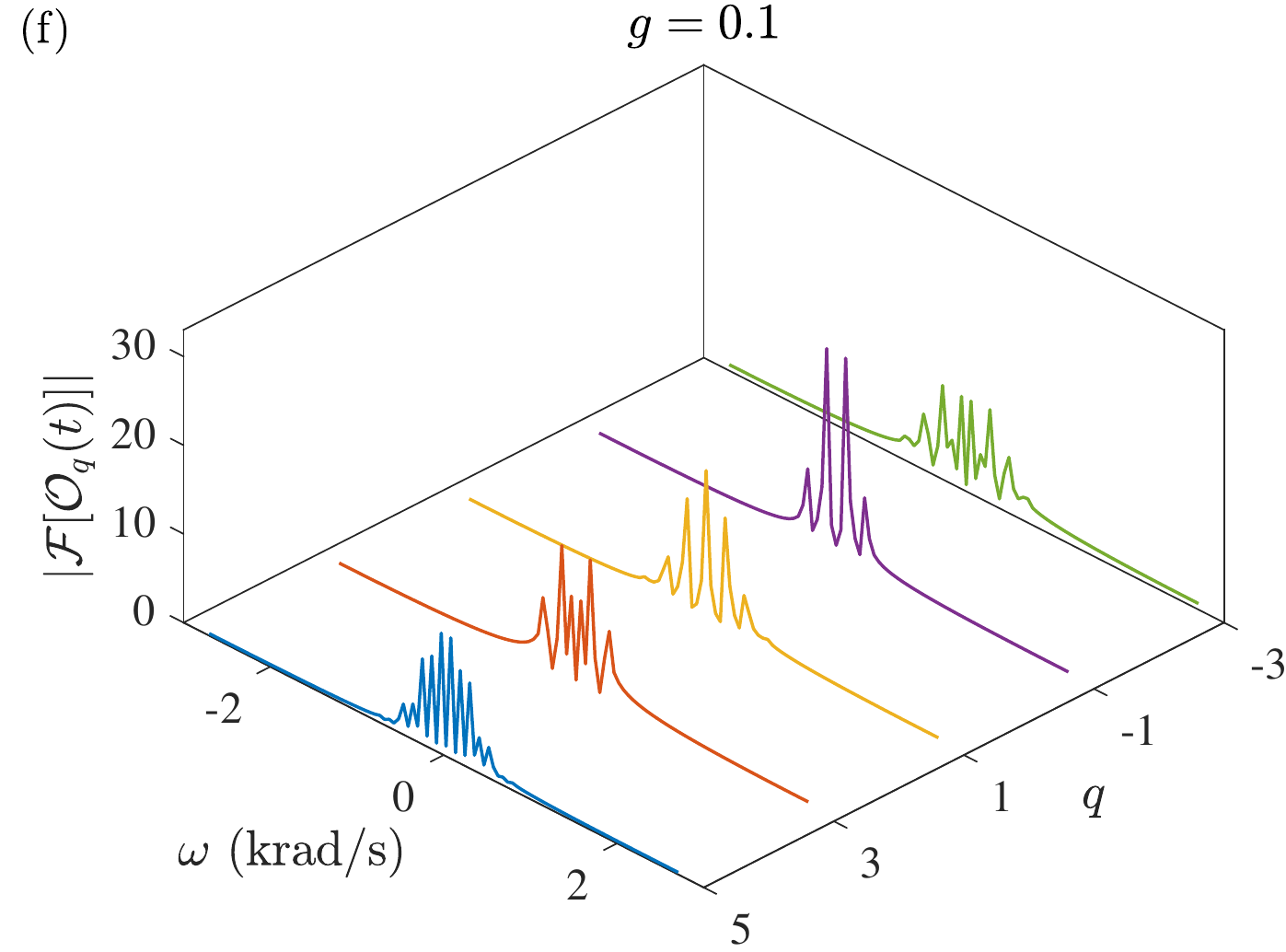}
}\\
	\subfigure{
		\includegraphics[trim = 3cm 9.8cm 3.8cm 9.4cm, clip, width=7.2cm ]{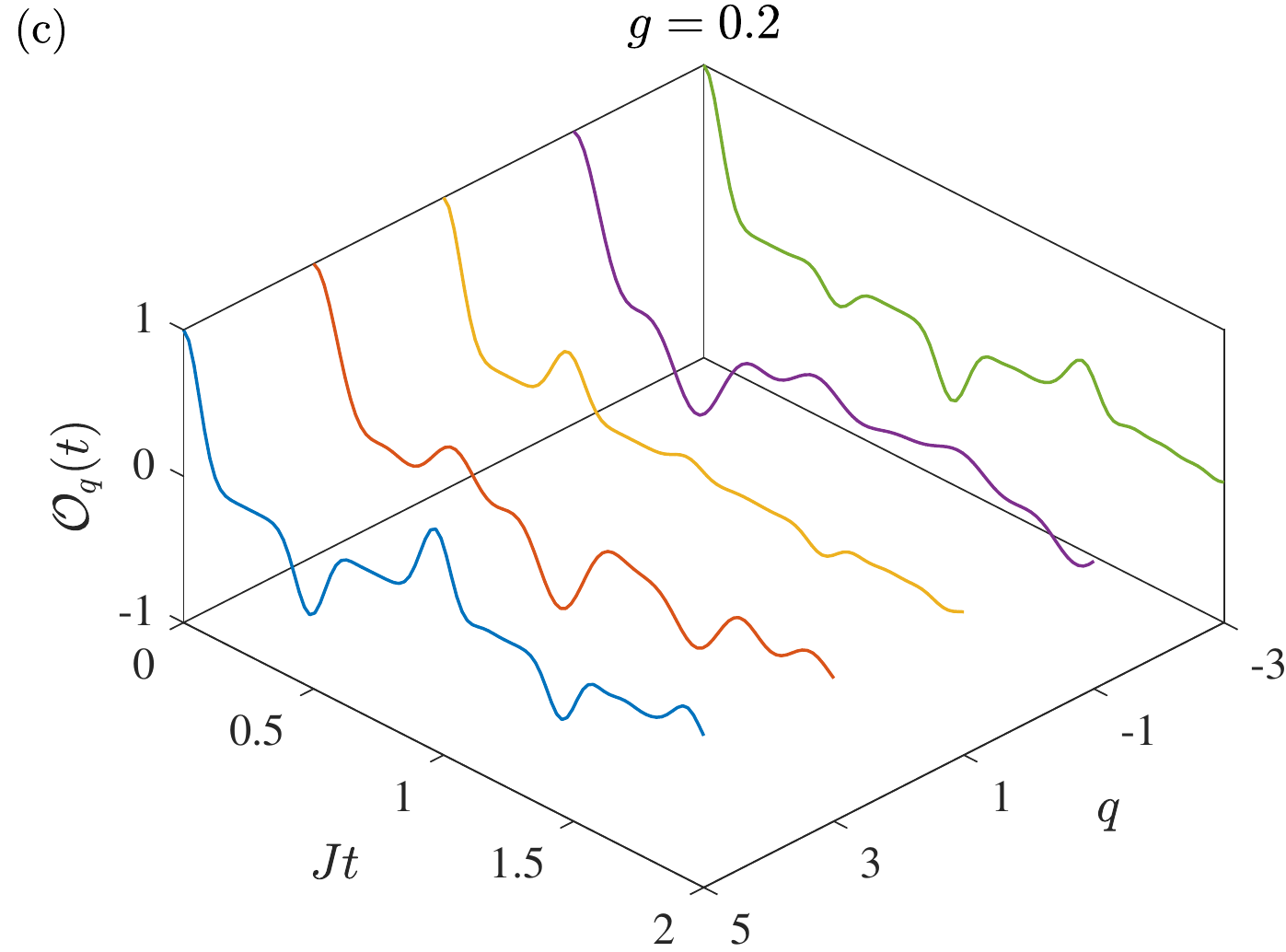}
	}\hspace{0.5cm}
\subfigure{
	\includegraphics[trim = 3cm 9.8cm 3.8cm 9.4cm, clip, width=7.2cm ]{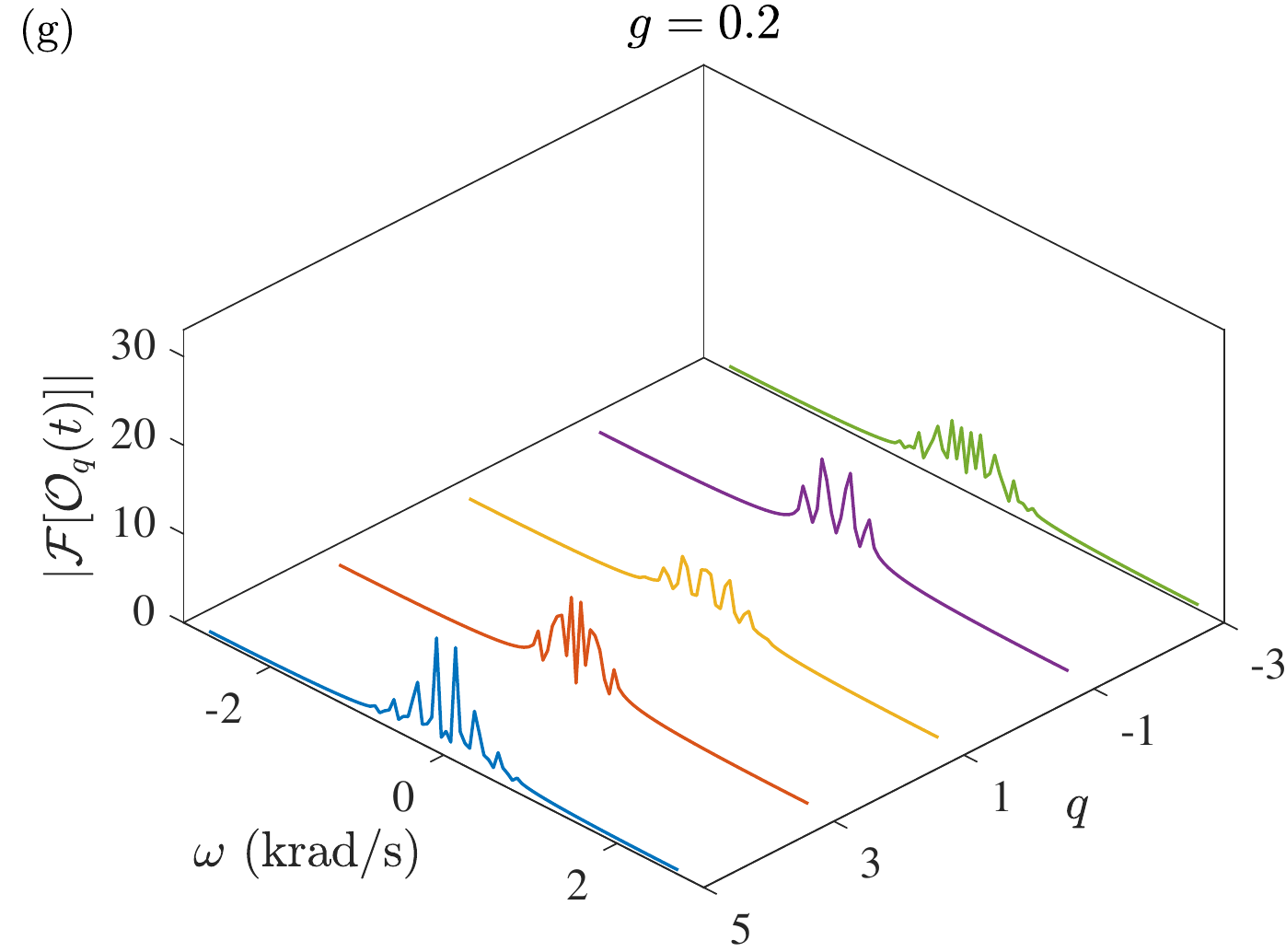}
}\\
	\subfigure{
		\includegraphics[trim = 3cm 9.8cm 3.8cm 9.4cm, clip, width=7.2cm ]{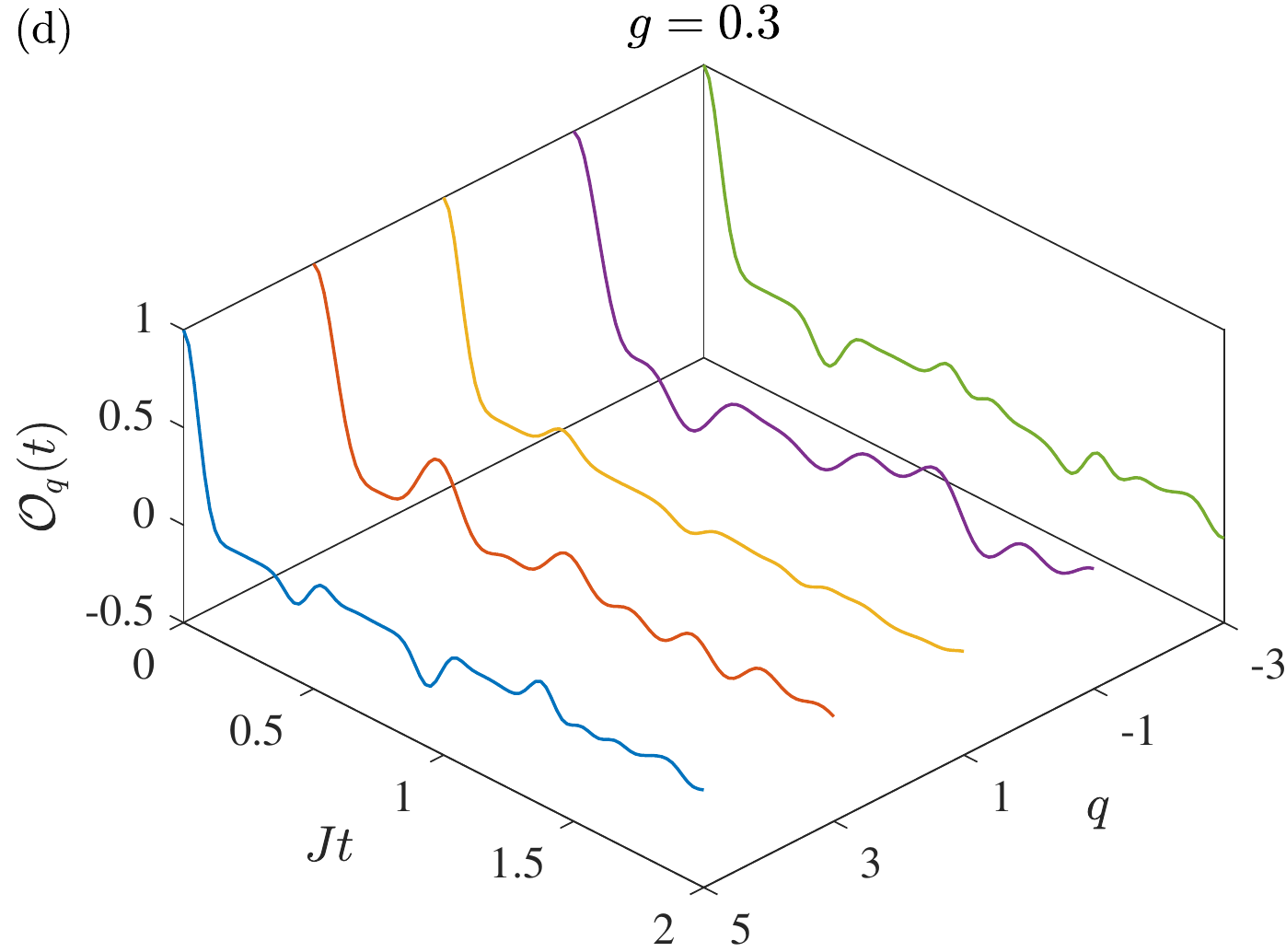}
	}
	\hspace{0.5cm}
	\subfigure{
		\includegraphics[trim = 3cm 9.8cm 3.8cm 9.4cm, clip, width=7.2cm ]{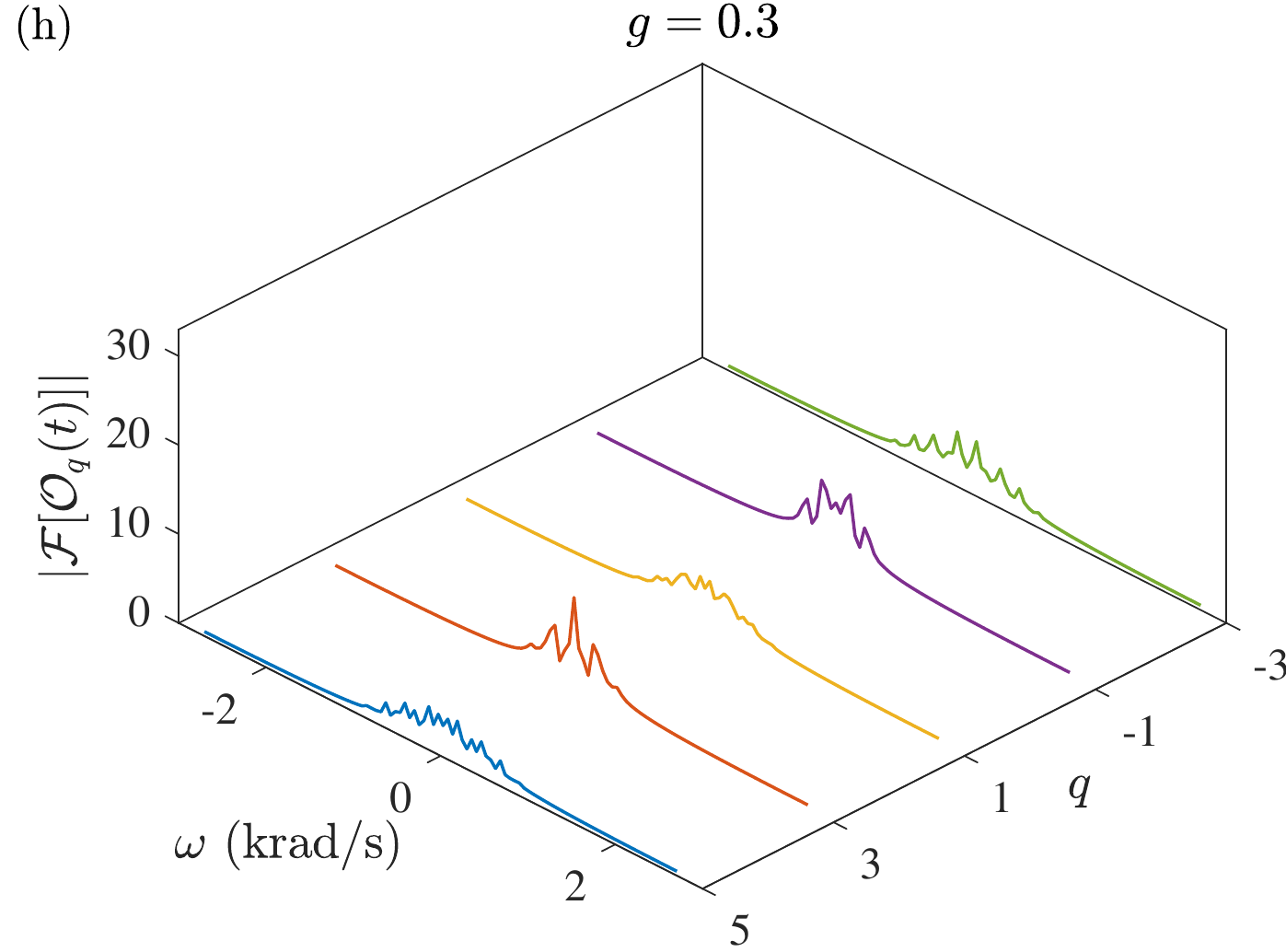}
	}
	\caption{Simulated time evolutions of OTOC functions for combination MQCs of coherence orders $q \in \{-3,-1,1,3,5\}$ for (a) $g=0$, (b) $g=0.1$, (c) $g=0.2$, and (d) $g=0.3$. Corresponding Fourier transforms $\mathcal{F}[\mathcal{O}(t)]$ are shown in (e-h). All the simulations are carried out for a two-branch HSTS ($K=2$).  Here no additional decoherence is introduced and all the decays are purely due to the information scrambling.}
	\label{simuotoc}
\end{figure*}
The impact of system size and decoherence for such as HSTS are discussed in Appendix \ref{appB}. 

Now we describe the preparation of combination MQCs between the central qubit and the first layer. Suppose the central spin $^{31}$P is initialized in $\ket{\pm}^P = (\ket{0}^P \pm \ket{1}^P)/\sqrt{2}$ and the surrounding $^1$H spins are in the state
\begin{equation}
\ket{\xi_n^{N_1}} = \ket{N_1-n,n}^H
\end{equation}
indicating $N_1-n$ spins in $\ket{0}$ state and $n \in [0,N_1]$ spins in $\ket{1}$ state.
We now apply a CNOT gate 
\begin{equation}
U_c = \left\{(\proj{0})^P \otimes \mathbbm{1}^H + (\proj{1})^P \otimes H_x\right\}\otimes \mathbbm{1}^F = U^\dagger_c
\end{equation}
with central $^{31}$P spin as the control and surrounding $^1$H spins as target.
The resulting state is
\begin{equation}
\rho_q^\pm(0) =  \proj{\psi_q^\pm} \otimes \mathbbm{1}^{F}/2^{N_2}
\label{rhoq}
\end{equation}
with
\begin{equation}
\ket{\psi_q^\pm } = \frac{
\ket{0}^P \ket{\xi_n^{N_1}} \pm
\ket{1}^P \ket{\xi_{N_1-n}^{N_1}}
}{\sqrt{2}},
\end{equation}
which represents a combination MQC with quantum number 
\begin{equation}
q = N_1-2n+1.
\end{equation}

In the following section, we describe scrambling of information out of these MQCs.

\section{Scrambling dynamics of combination MQCs}
Following the discussion preceding Eq. \ref{rhodelta}, we choose 
%
\begin{equation}
\rho_\Delta(0) = \rho^x_q(0) = \rho^+_q(0) -\rho^-_q(0).
\label{rhox}
\end{equation}
Therefore, we consider the unambiguous study of OTOC dynamics with the following operators:
\begin{align}
A(0) & = \rho_q^x(0) ~~\mbox{and}
\nonumber \\
B(t) & = U^\dagger(t).
\end{align}

In this case, the OTOC function becomes
\begin{align}
{\cal O}_q(t)  \approx &
\mathrm{Re}\left[\langle B^\dagger(t)A^\dagger(0)B(t)A(0)\rangle_{\beta=0} \right]
\nonumber \\
&=\mathrm{Tr}\{U(t)\rho_q^x(0)U^\dagger(t) ~\rho_q^x(0)\} 
\nonumber \\
&= \mathrm{Tr}\{\rho(t)\rho_q^x(0)\} .
\label{fqt} 
\end{align} 
The propagator $U(t)$ involves all the spins including those in the second layer and may lead to an effective leakage of coherence from the initial $q$-quantum combination MQC subspace.

\subsection{Numerical simulations}

We have performed the following numerical simulations to gain more insight into the scrambling dynamics of combination MQCs. Considering the computational cost, we simulated only the partial system with $K = 2$ in the Hamiltonian given in Eq. \ref{Hamil}.  Here no decoherence is introduced, and the observed effects are only due to the scrambling dynamics.

 Fig. \ref{simuotoc} displays the simulated OTOC for various coherence orders $q$ and $Jt$ for various $g$ values. 
For the case $g = 0$, the dynamics is integrable, as shown in Fig. \ref{simuotoc}(a). In this case, the OTOC function shows periodic oscillations for all the MQCs, without any effective decay, suggesting no information scrambling.
Note that the profiles of $q=5$ and $q=-3$ match exactly.  This is because, the  corresponding states $\ket{\psi_5}$ and $\ket{\psi_{-3}}$ differ only by the state of the central qubit which does not evolve under the scrambling Hamiltonian in the absence of the external fields.  Similarly, $q=3$ and $q=-1$ also match for the same reason.

However, once the external fields are applied, i.e., $g>0$, the dynamics becomes non-integrable. In this case, the OTOC oscillations become nonperiodic, as shown in Figs. \ref{simuotoc}(b-d). More importantly, the OTOC profiles now suffer from effective decays due to a gradual loss of information out of the MQC $\rho_q^x (0)$.  In fact, the stronger the strength $g$ of the external fields, the more efficient is the scrambling.  This dependence of scrambling with nonintegrability of dynamics has also been noted earlier \cite{li2017measuring} in the context of a spin chain. 

Further insight can be obtained by looking at the frequency profiles of OTOC functions.  Fig. \ref{simuotoc}(e-h) display Fourier transforms  ${\cal F}({\cal O}_q(t))$ for various combination MQCs at different $g$ values.  At $g=0$, the spectral lines are sharp, indicating finite frequency components.  However, as we introduce the external fields, i.e., for $g > 0$, we find the emergence of more frequency components, which indicates a stronger leakage of information leading to more efficient scrambling.  As $g$ increases further, we observe an effective smoothening of frequency profiles. At this point, the time domain decay profiles appear almost exponential decays, and therefore, it  becomes hard to differentiate them from decoherence induced decays. This fact emphasizes the importance of CTP in  practical situations.
In the next subsection, we experimentally apply  CTP to reveal information scrambling for filtered combination quantum coherence $q = -1$. 

\subsection{Experiments}
\begin{figure}[H]
	\subfigure{\includegraphics[trim = 4cm 8.8cm 4cm 9cm, clip, width=8.5cm ]{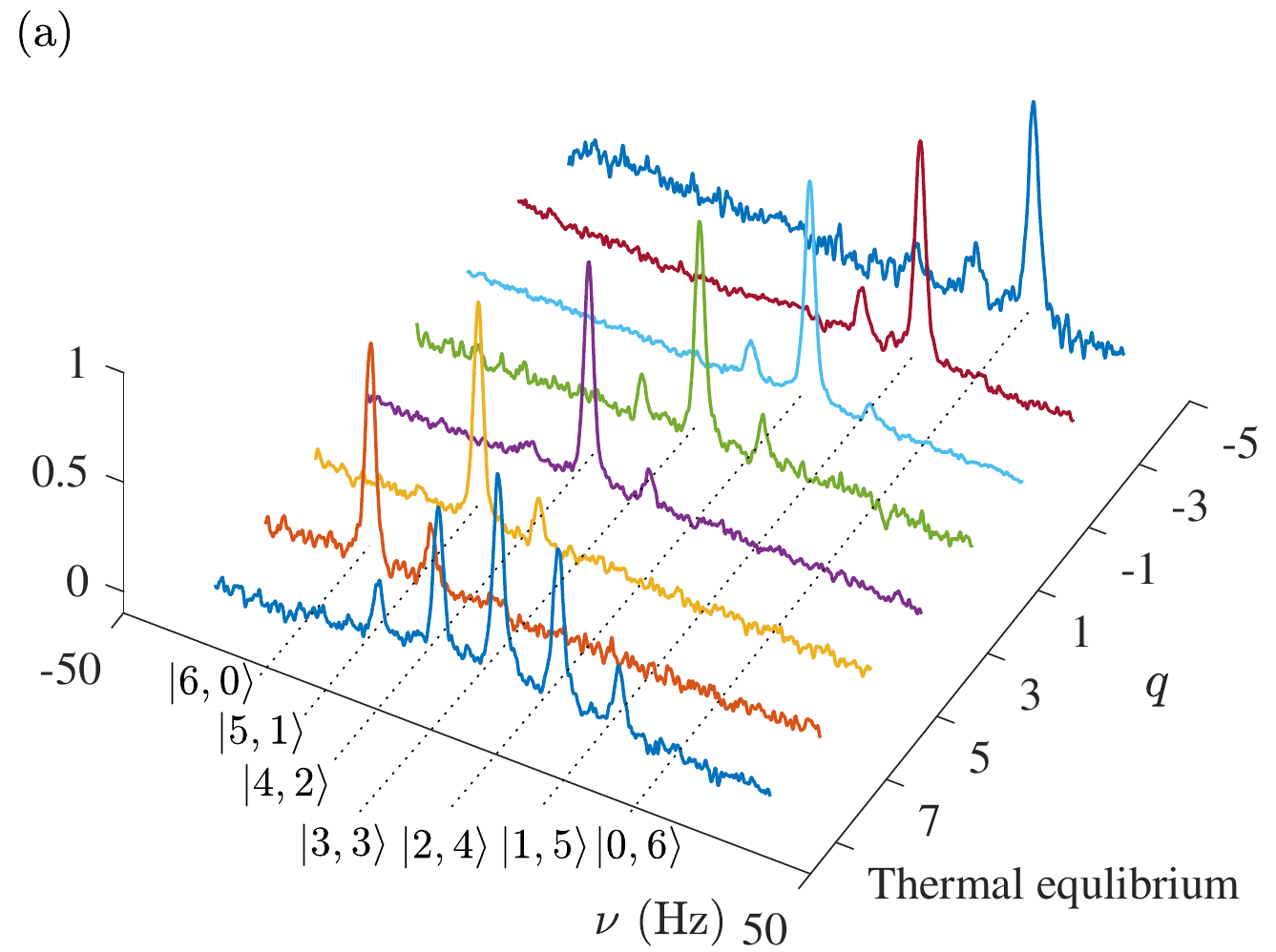}}
	\subfigure{\includegraphics [trim = 0cm 6.8cm 1cm 6.8cm, clip, width=8.5cm ]{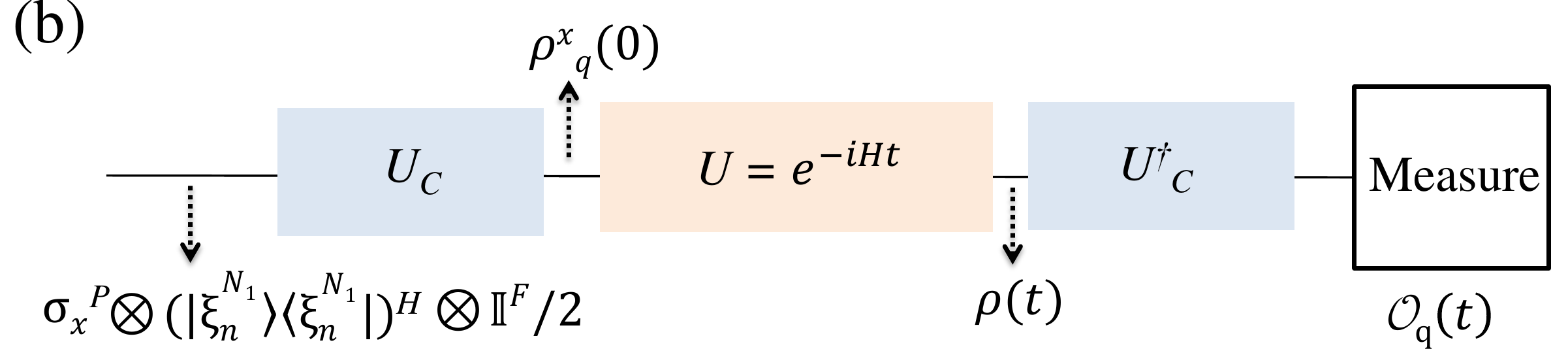}}\\\vspace{0.5cm}
	\subfigure{\includegraphics[trim = 0cm 1.3cm 1cm 0cm, clip, width=8.5cm ]{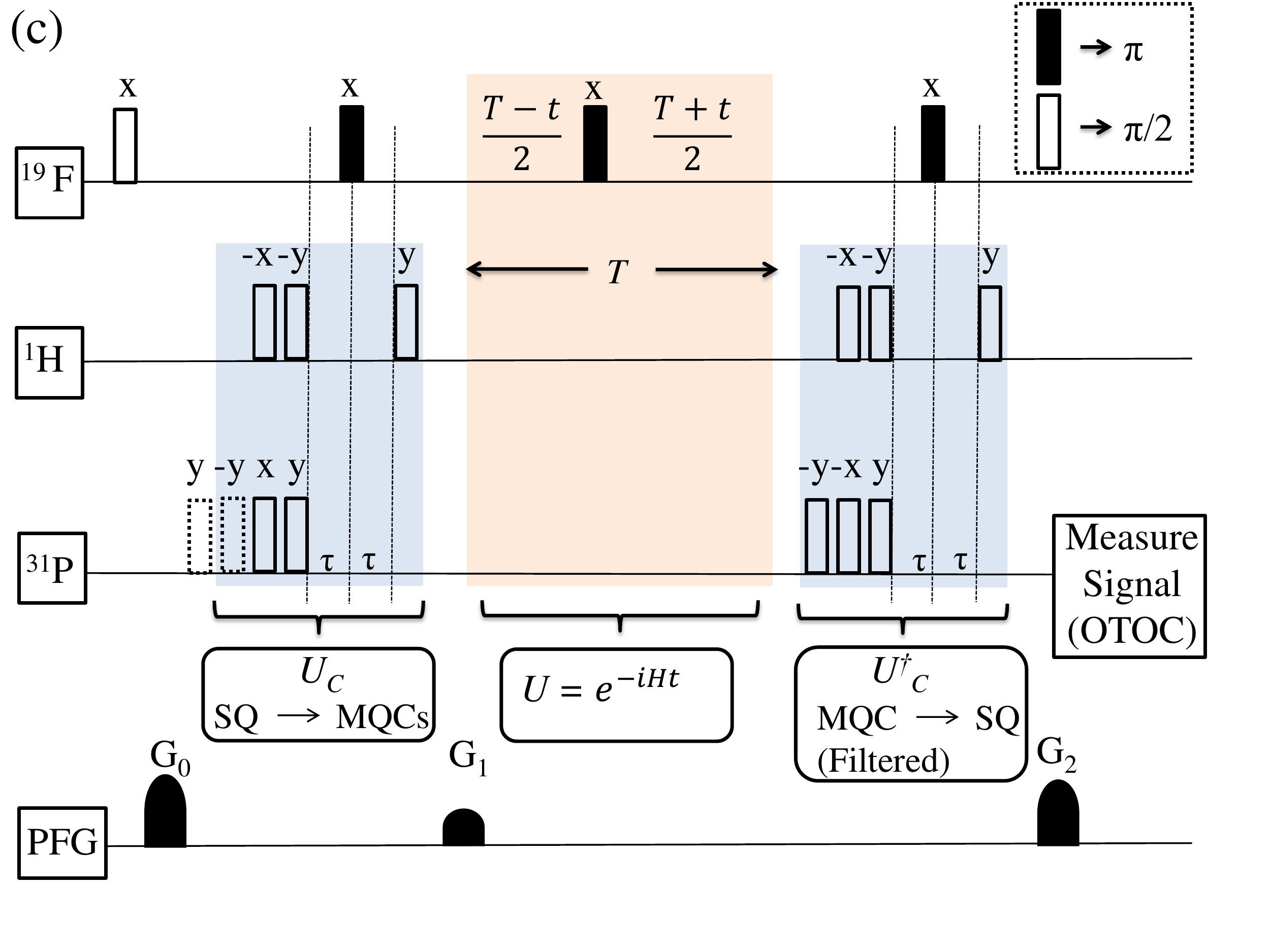}}
	
	\caption{(a) A reference $^{31}$P NMR spectrum (first trace) and spectra corresponding to various quantum numbers $q$ as indicated. The transitions are labeled by corresponding  states  $\ket{\xi_n^{N_1}}$ of $^1$H spins. (b) Schematic illustration of the experimental protocol to study scrambling dynamics of MQCs. (c) The NMR pulse sequence to implement the protocol in (b). The open and filled rectangles correspond to $\pi/2$ and $\pi$ rotation with phases as indicated. Here PFG denotes pulse field gradients along $z$-direction used to select a particular coherence pathway between MQCs and SQ (single quantum coherence) and $\tau = 1/4J$ indicates evolution time under coupling Hamiltonian given in Eq. \ref{internal}.}
	\label{circuit}
\end{figure}

The NMR experiments were carried out in a  Bruker NMR spectrometer with a static field of $11.2$ T.  As described in section III, the sample consisted of  tris(2,2,2-trifluoroethyl) phosphite (see Fig. \ref{mol}) dissolved in deuterated dimethyl sulphoxide (0.05 ml in 0.5 ml).  The sample was maintained  at an ambient temperature of 298 K. The $^{31}$P NMR spectra corresponding various filtered MQCs along with a reference spectrum are shown in Fig. \ref{circuit}(a). Each transition is labeled by spin states $\ket{\xi_n^{N_1}}$ of the $^1$H spins.  

The experimental protocol is described schematically in Fig. \ref{circuit} (b). Starting from thermal equilibrium, we prepare $\rho_q^x (0)$ (see Eq. \ref{rhox}) using a $(\pi/2)_y$ pulse on $^{31}$P followed by a CNOT gate $U_c$. Note that the CNOT gate is applied in parallel to all the $^1$H spins exploiting the star-topology of the system.  Then we use the CTP method to control the scrambling time $t$ with fixed total time $T$ as described in Fig. \ref{ctp}.  The final state $\rho(t)$ is converted into the observable single quantum magnetization of the central spin using a second CNOT gate $U^\dagger_c = U_c$.
The resulting signal is
\begin{align}
s_q^x(t) &= \mathrm{Tr} \left[
(\sigma_{x}^P \otimes \proj{\xi^{N_1}_n} \otimes \mathbbm{1}^F ) ~ U^\dagger_c \rho(t)  U_c \right]
\nonumber \\
 &= \mathrm{Tr} \left[
U_c (\sigma_{x}^P \otimes \proj{\xi^{N_1}_n} \otimes \mathbbm{1}^F ) U^\dagger_c ~  \rho(t)   \right]
 \nonumber \\
 &= \mathrm{Tr} \left[
 U_c \left\{  \ket{+}^P \bra{+}^P\otimes \proj{\xi^{N_1}_n} \otimes \mathbbm{1}^F \right\}  U^\dagger_c  ~ \rho(t)  \right]
 \nonumber \\
 &~~- \mathrm{Tr} \left[
 U_c \left\{  \ket{-}^P \bra{-}^P\otimes \proj{\xi^{N_1}_n} \otimes \mathbbm{1}^F \right\}  U^\dagger_c  ~ \rho(t)  \right]
 \nonumber \\
 &= \mathrm{Tr} \left[
 \rho_q^+(0) \rho(t)  \right]
 - \mathrm{Tr} \left[
\rho_q^-(0) \rho(t)  \right]
 \nonumber \\
 &= \mathrm{Tr}\left[\rho_q^x(0) \rho(t) \right]
 \nonumber \\
& = {\cal O}_q(t),
\end{align}
where we have used Eq. \ref{rhoq} and \ref{rhox}.
Thus OTOC can be directly extracted from the NMR signal $s_q^x(t)$ of the central $^{31}$P spin.

   The NMR pulse sequence for the preparation of the combination MQCs is shown in Fig. \ref{circuit} (c). We start with the application of a $(\pi/2)_x$ pulse on $^{19}$F spins followed by a pulsed-field-gradient (PFG) $G_0$ along $z$ direction  to prepare them in the maximally mixed $(\mathbbm{1}/2)^{\otimes N_2}$ state. Subsequently a Hadamard gate using a $(\pi/2)_y$ is applied on $^{31}$P. The CNOT operation is realized via $^{1}$H-$^{31}$P J-coupling.  During this period we refocus the interactions with $^{19}$F spins using a $(\pi)_x$ pulse on  $^{19}$F. The unambiguous study of scrambling is carried out using the CTP method as described in the previous section (see Eq. \ref{cpt}).  The time-reversal step in CTP is also realized using a $(\pi)_x$ pulse on  $^{19}$F.  We vary the time parameter $t$ by holding the total time $T$ constant, so that the decoherence effects are same in all the experiments, while scrambling duration is systematically varied.
  Finally, MQCs are converted back to an observable single-quantum coherence.  A specific MQC $\rho^x_q (0)$ of a particular quantum number $q$ is filtered out by a pair of PFGs of strengths $G_1$ and $G_2$ as shown in Fig. \ref{circuit} (c).  The ratio of the PFGs to filter the $q$-quantum combination MQC is set to \cite{shuklaNOON}
  \begin{equation}
  	\frac{G_1}{G_2} = -\frac{\gamma_P+(q-1)\gamma_H}{\gamma_P },
  \end{equation}
where $\gamma_P$ and $\gamma_H$ are gyro-magnetic ratios of $^{31}$P and $^{1}$H respectively.

\begin{figure*}
	
	\subfigure{
		\includegraphics[trim = 3.1cm 8cm 3.3cm 8.5cm, clip, width=8.2cm ]{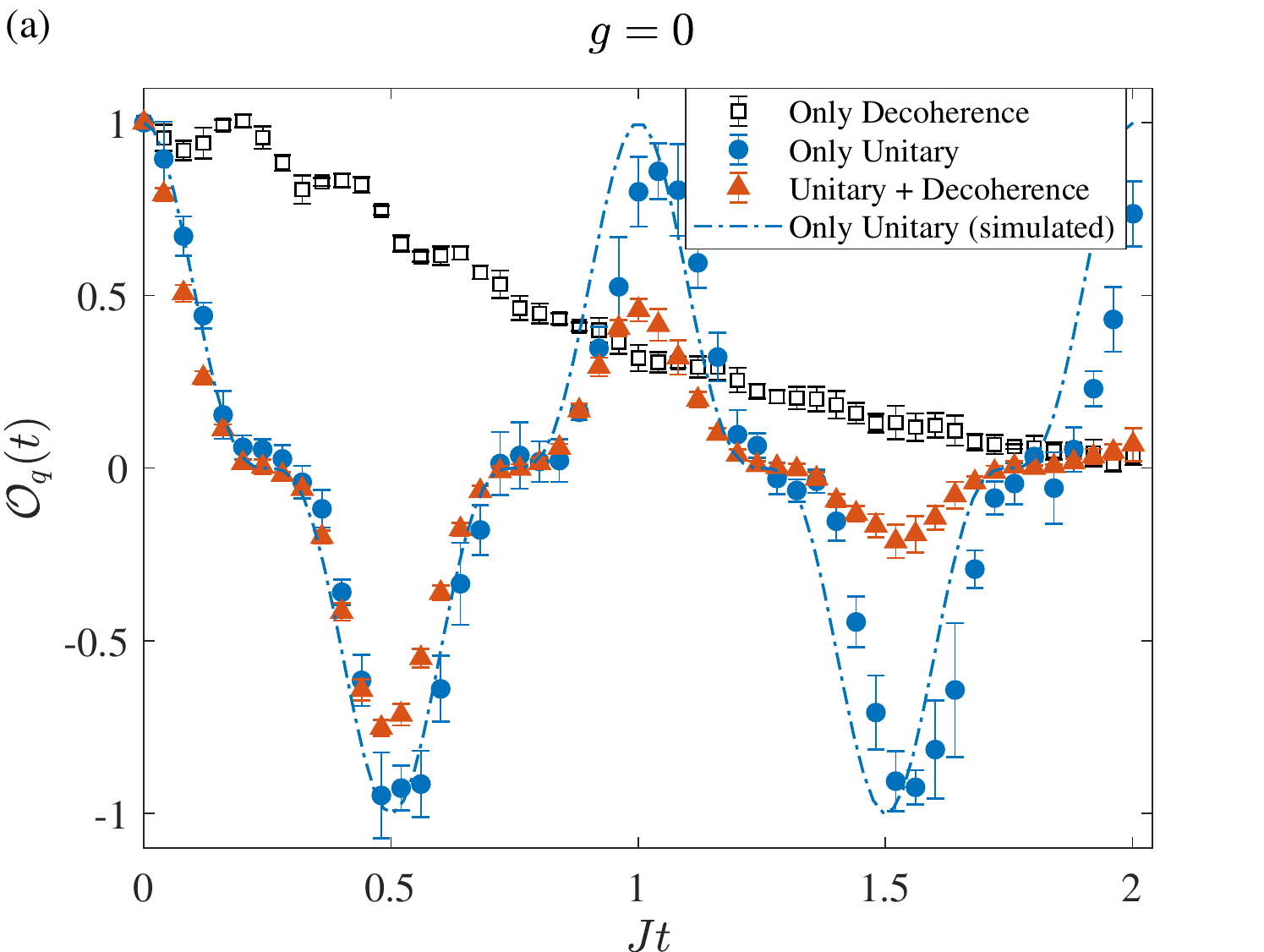}
	}\hspace{0.5cm}
	\subfigure{
		\includegraphics[trim = 3.1cm 8cm 3.3cm 8.5cm, clip, width=8.2cm ]{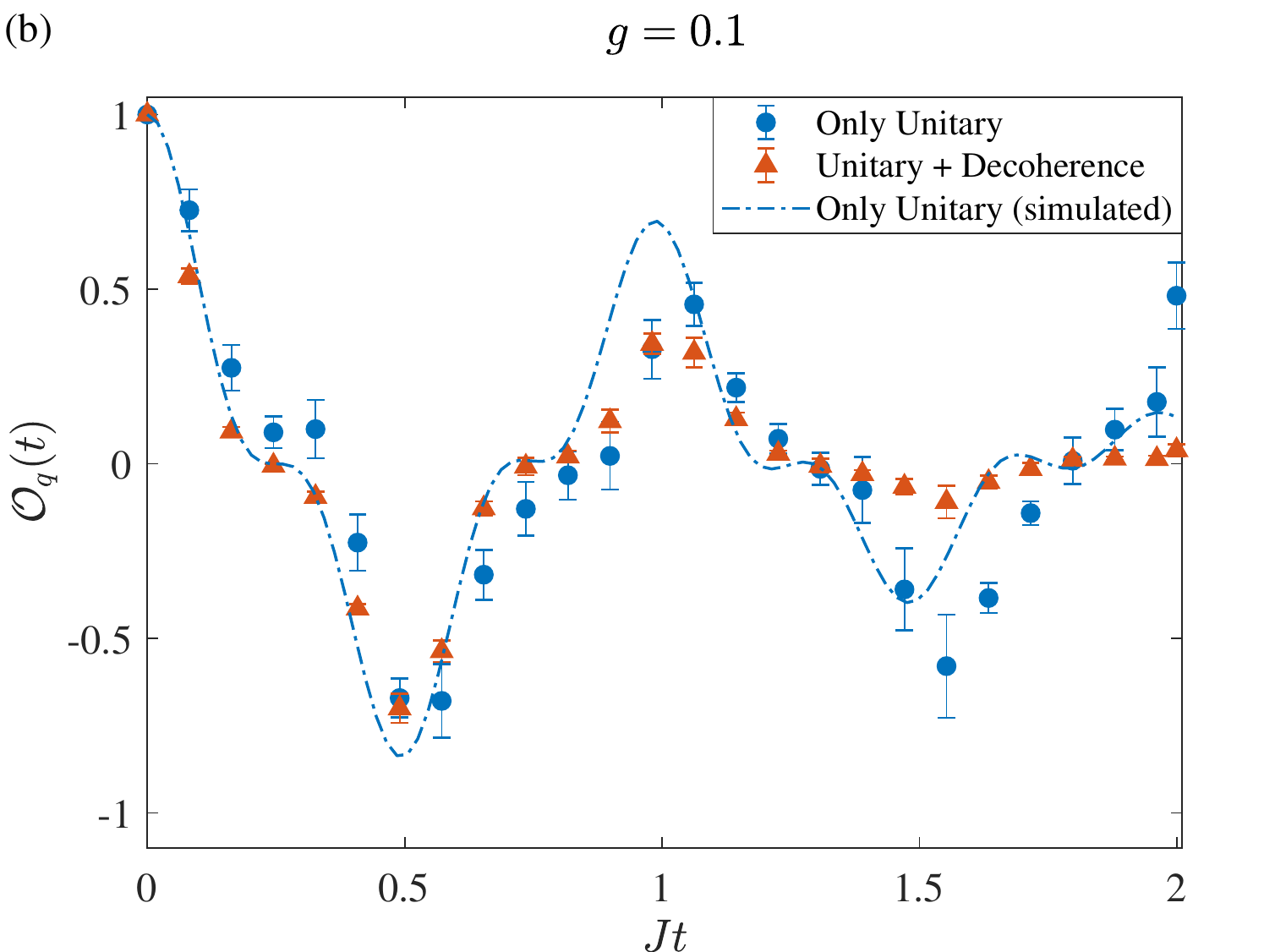}
	}
	\subfigure{
		\includegraphics[trim = 3.1cm 8cm 3.3cm 8.5cm, clip, width=8.2cm]{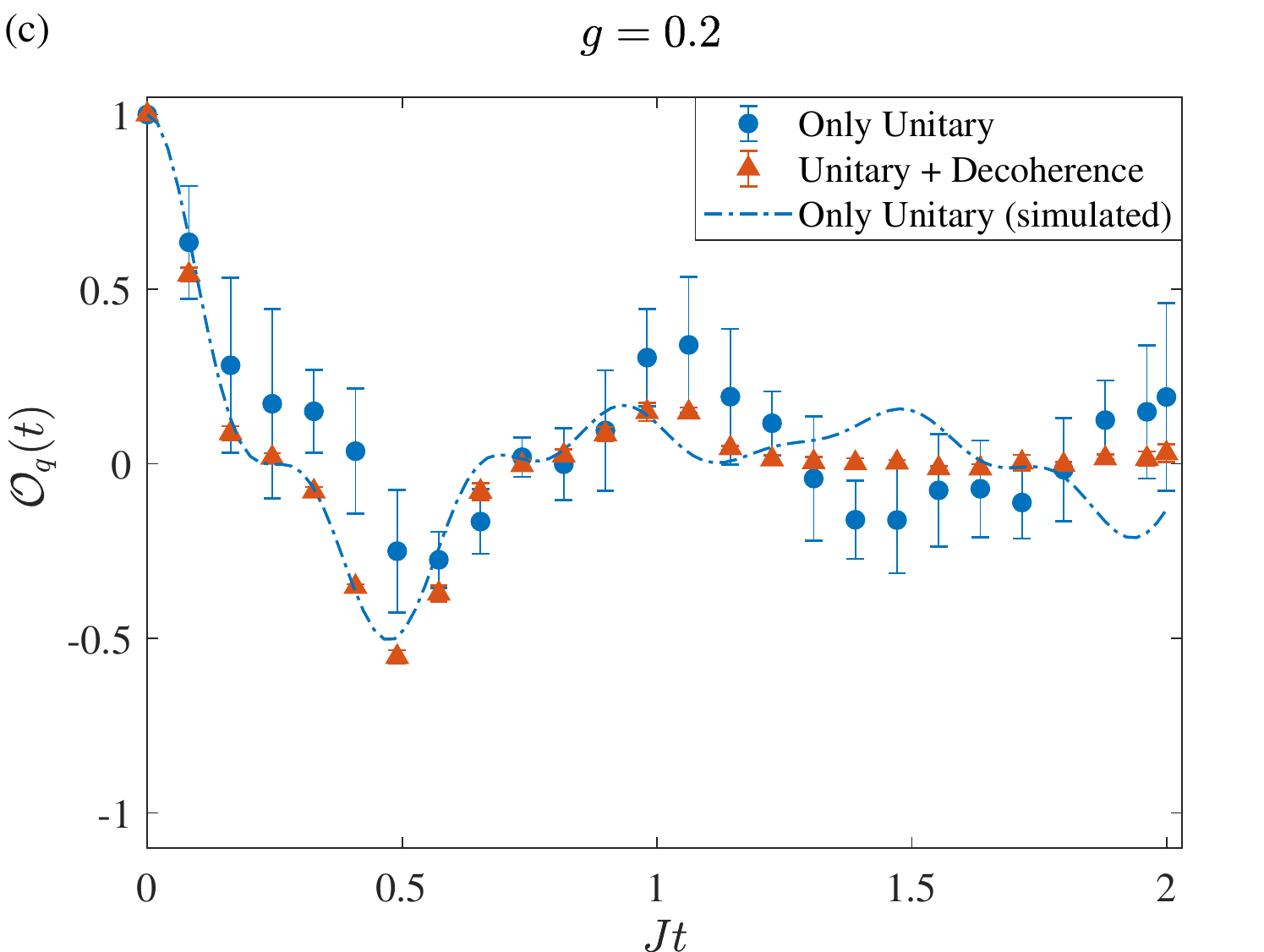}
	}\hspace{0.5cm}
	\subfigure{	\includegraphics[trim = 3.1cm 7.5cm 4cm 8.5cm, clip, width=8.2cm ]{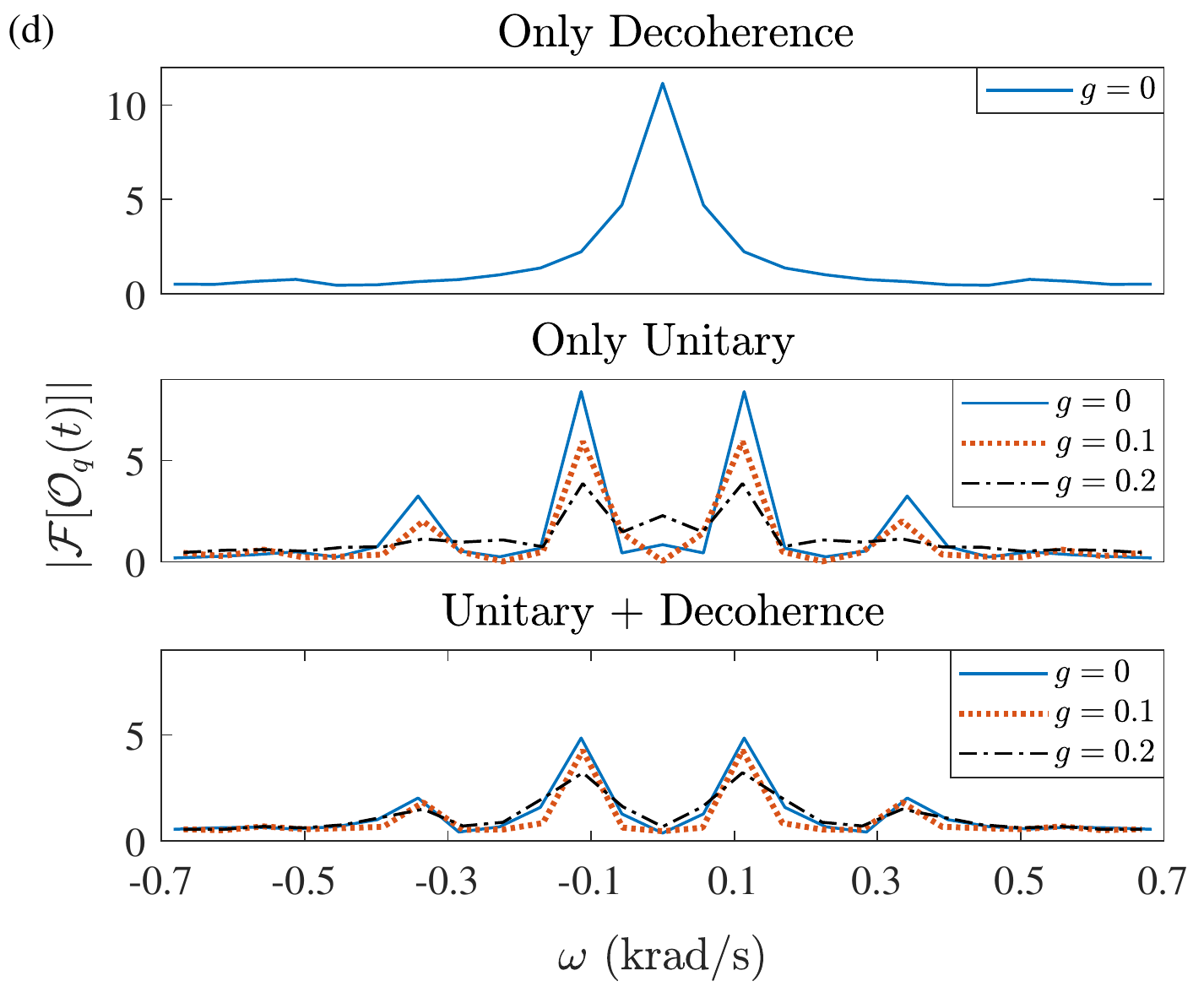}}
	\caption{Experimentally measured OTOC corresponding to time evolution of $q = -1$ quantum coherence in (a) integrable ($g = 0$) and (b-c) non-integrable ($g \neq 0$) regime. Dashed lines in (a-c) are obtained by numerical simulation using two-branch ($K = 2$) HSTS. Corresponding Fourier transform profiles are shown in (d).}
	\label{int_non_int}
\end{figure*}

Figs. \ref{int_non_int} (a-c) display the experimentally measured OTOC functions corresponding to the quantum number $q=-1$  for various values of the nonintegrability parameter $g$.  
We have chosen $q = -1$ coherence because of its comparatively longer coherence time than the other MQCs.
Fig. \ref{int_non_int}(a) displays OTOC evolution for $g=0$ which belongs to the integrable regime and hence does not introduce scrambling.  Here we are able to separate all three types of dynamics as follows:
\begin{itemize}
\item[(i)]  Only decoherence (without unitary evolution $U(t)$): It is realized by effectively nullifying the interaction between $^1$H spins of the first layer and $^{19}$F spins of the second layer using a $(\pi)_x$ pulse on $^{19}$F spins at the center of the time evolution. The decay profile leads to an effective coherence time $T_2^*\simeq140$ ms (empty squares in Fig. \ref{int_non_int}(a)).  
\item[(ii)] Unitary dynamics along with decoherence: It is realized by allowing the interaction of $^1$H spins with $^{19}$F spins (triangles in Fig. \ref{int_non_int}(a)). This dynamics shows an oscillatory decay of OTOC, which in practical timescales of observation can be confused with the scrambling.  
\item[(iii)] Pure unitary dynamics - realized by CTP (circles in Fig. \ref{int_non_int}(a)).  Here we observe almost decay-less oscillations with strong revivals of OTOC confirming the absence of genuine scrambling.  The higher error bars, in this case, are due to lower signal to noise ratio.
\end{itemize}

Since case (ii) is the combined effect  of the case (i) and (iii), one may expect the curve with triangles to match with the product of  curves with squares and circles. However,  here we find an interesting observation: the triangles over-shoot the empty squares at certain time instants (e.g. near $Jt = 1$) possibly signaling an information back-flow due to non-Markovianity \cite{breuer2016colloquium,de2017dynamics}.

In Fig. \ref{int_non_int}(b) and (c)  we show time evolution of OTOC with $g= 0.1$ and $g=0.2$ respectively.  Here the presence of external fields leads to nonintegrable dynamics and consequently exhibit information scrambling.  As a result, the OTOC does not show revivals back to the initial value.  One can compare the OTOC data with unitary + decoherence in (a) (triangles) with OTOC data with only unitary (circles) in (c).  While both 
show decaying revivals, the former is devoid of scrambling while the latter is purely due to scrambling.  This suggests the importance of separating the decoherence effects before quantifying scrambling.  Fig. \ref{int_non_int}(d) displays the Fourier transform of the OTOC data.  As discussed in the previous subsection, we find broader and more dispersed spectral lines as we increase $g$, indicating stronger information scrambling.

\section{Conclusions}
In this work, we have studied scrambling of information in a double layered star-topology system. This topology  allows us to efficiently prepare multiple quantum coherences involving central qubit and the first layer qubits. The scrambling is introduced in a controlled manner using the tunable external fields. 

A major hurdle in the unambiguous study of scrambling is to account for the contribution from decoherence to OTOC dynamics.   In this regard, we proposed a constant-time protocol which enables us to filter out contribution  solely from scrambling. 

Using a sixteen-spin double layered star-topology NMR system, we experimentally demonstrated the unambiguous  study of scrambling of  information stored in   
the combination multiple quantum coherence involving central qubit and six satellite qubits in the first layer.
With the help of  constant time protocol, we could clearly separate decoherence effects and obtained OTOC profiles exclusively characterizing scrambling effects.  While we observed signatures of non-Markovian evolutions, it calls for  further detailed investigation in this direction.

Although the brute-force simulation of the complete system was computationally too expensive, it was nevertheless easier to tune the external field, control the scrambling rate and measure the OTOCs in the NMR spectrometer.  In a way, it is a demonstration of the supremacy of quantum simulations over the classical analogs.
Therefore, we expect to see more applications of such  star-topology systems in studying many-body phenomena because of convenient manipulation  allowed by higher symmetry.
  
\section*{Acknowledgments}
This work was supported by DST/SJF/PSA-03/2012-13 and CSIR 03(1345)/16/EMR-II.
  
\appendix
\section{OTOC for highly mixed single-qubit initial state }
\label{appA}
Eq. \ref{rhodelta} can be derived as follows
\begin{align}
{\cal O}(t)  &= \mathrm{Re}[\langle B^{\dagger}(t) \rho^\dagger(0) B(t) \rho(0)\rangle_{\beta \rightarrow 0}] \nonumber \\
 &= \mathrm{Re}[\mathrm{Tr}\{ B^{\dagger}(t) \rho^\dagger(0) B(t) (\rho(0))^2\}] \nonumber \\
 &=\mathrm{Re}\left[\mathrm{Tr}\left\{ B^{\dagger}(t) \left(\mathbbm{1}/2 + \epsilon \rho_\Delta (0)\right) B(t) \left(\mathbbm{1}/2 + \epsilon \rho_\Delta (0)\right)^2\right\}\right] 
 \label{a1} \nonumber
\end{align}
The right hand side produces following six terms
\begin{align}
&\frac{1}{8} \mathrm{Re}[\mathrm{Tr}\{B^\dagger(t) B(t)\}] \rightarrow \frac{1}{4} \nonumber\\
&\frac{\epsilon}{2} \mathrm{Re}[\mathrm{Tr}\{B^\dagger(t) B(t) \rho_\Delta(0)\}] \rightarrow 0 \nonumber\\
&\frac{\epsilon^2}{2} \mathrm{Re}[\mathrm{Tr}\{B^\dagger(t) B(t) (\rho_\Delta (0))^2\}] \rightarrow \frac{\epsilon^2}{8} \mathrm{Tr}\{(\rho_\Delta(0))^2\} \nonumber\\
&\frac{\epsilon}{4} \mathrm{Re}[\mathrm{Tr}\{B^\dagger(t) \rho_\Delta(0) B(t)\} ] \rightarrow 0 \nonumber\\
&\epsilon^2 \mathrm{Re}[\mathrm{Tr}\{B^\dagger(t) \rho_\Delta(0) B(t) \rho_\Delta(0)\} ] 
\rightarrow *
\nonumber\\
&\epsilon^3 \mathrm{Re}[\mathrm{Tr}\{B^\dagger(t) \rho_\Delta(0) B(t) (\rho_\Delta (0))^2\} ] \rightarrow \mathrm{negligible~for~}\epsilon \ll 1 \nonumber\\
\nonumber
\end{align}
As clear from the above, only the term indicated by * has information about the OTOC dynamics.  
Plugging these values back, we get
\begin{align}
{\cal O}(t) &= \frac{1}{4} + \frac{\epsilon^2}{8}\mathrm{Tr}\{(\rho_\Delta(0))^2\}\nonumber\\ 
&~~+ \epsilon^2 \mathrm{Re}[\mathrm{Tr}B^\dagger(t) \rho_\Delta(0) B(t) \rho_\Delta(0) ],
\end{align}
Hence
\begin{equation}
{\cal{O}} (t) \sim \mathrm{Re}[\mathrm{Tr}\{B^{\dagger}(t) \rho_\Delta(0) B(t) \rho_\Delta(0)\}],
\end{equation}
up to $\epsilon^2$ factor and a constant background.  It is interesting to note that, in NMR conditions, other measures of quantum correlations, such as Quantum Discord \cite{katiyar2012evolution} and deviations in von Neumann entropy \cite{krithika2019nmr}, are also measured in units of $\epsilon^2$.

\section{Impact of system size and decoherence}
\label{appB}
It is useful to have some idea on how the extent of scrambling scales with the size of  HSTS.  In this regard, we consider the scrambling of information initially localized on the central spin (Fig. \ref{mol}) onto the $N_2$ spins in the second layer via $N_1$ spins in the first layer.  To this end, we choose 
$$A(0) = \sigma_y^P ~~~\mbox{and}~~~ B(t) =  U^\dagger(t) S_y^F U(t).$$ 

\begin{figure}
	
\includegraphics[trim = 3.2cm 9.4cm 3.8cm 9.4cm, clip, width=8.5cm ]{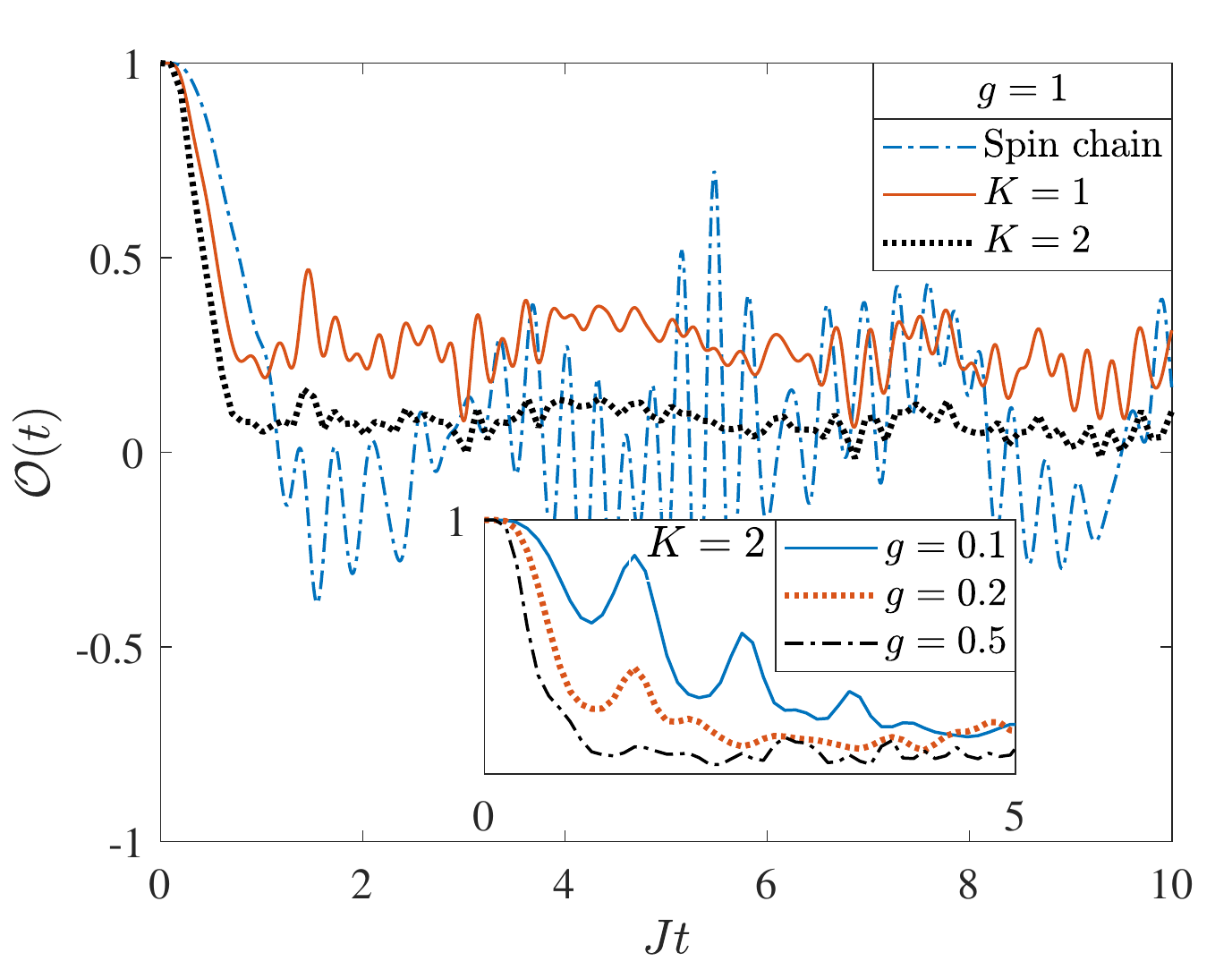}
	\caption{Dynamics of OTOC for the experimental system shown in Fig \ref{mol}.  Here  $A(0) = \sigma_y^P$ and $B(0) = S_y^F$ with $[A(0), B(0)] = 0$.  The OTOC measures scrambling of information from the central qubit to the third layer. Though evolution in the presence of both $x$ and $z$ field is shown, Hamiltonian gives rise to scrambling even in the absence of $z$ field. In the inset, variation of OTOC with the nonintegrability parameter $g$.  }
	\label{impact}
\end{figure}

Since simulating the exact dynamics of the entire system with three branches consisting of 16 spins in Fig. \ref{mol} is computationally expensive, we limit ourselves to partial system sizes.
For the integrable regime, i.e., nonintegrability parameter $g = 0$, OTOC function remains uniformly unity since the commutator $C(t)$ vanishes at all times owing to the fact
that $B(t)$ can only develop multi-spin orders with protons in the first layer with which it is directly interacting.  Hence information remains localized within the first layer and never scrambled onto the second layer.  Even for small values of $g$, OTOC function deviates from unity and starts oscillating (see the inset of Fig. \ref{impact}).
Now we set $g = 1$ and look at the dependence of scrambling on system size  as shown in Fig. \ref{impact}.

\begin{figure}
	\includegraphics[trim = 3.4cm 9.4cm 4cm 9.8cm, clip, width=8.5cm ]{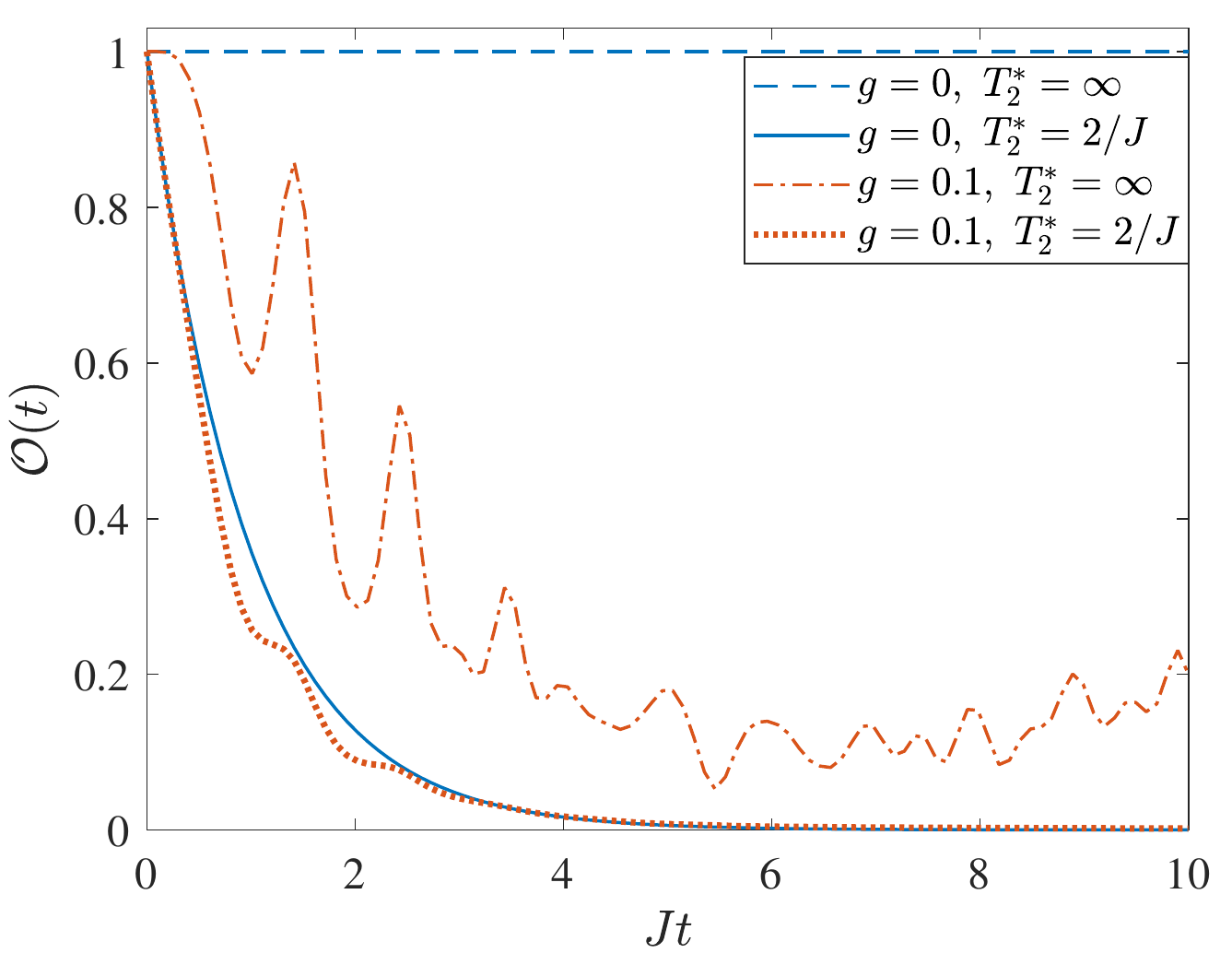}
	\caption{Ambiguity in the estimation of information scrambling due to the presence of decoherence simulated using two-branch HSTS.}
	\label{deco}
\end{figure}
We use Lindblad based approach to simulate the combined effects of scrambling and decoherence  with completely correlated dephasing model \cite{alonso2019out}. We introduce decoherence in the system by single-qubit dephasing modeled using the Lindblad equation
\begin{equation}
\frac{d\rho}{dt} = -i[H,\rho] + \gamma\sum_{i = 1}^{N}  \big(L_i \rho L_i^{\dagger}- \frac{1}{2} \{L_i^{\dagger} L_i,\rho\}\big),
\end{equation}
where  $N$ is number of spins, $L_i = \sigma_z^{i}$ and $\gamma = 1/(2T_{2}^*)$ is transverse relaxation  rate. To numerically simulate decoherence dynamics of OTOC, we update the density matrix in the following way
\begin{align}
\rho(t)\rightarrow & \gamma dt \sum_{i}  L_i U(dt)\rho(t)U^{\dagger}(dt)L^{\dagger}_i \nonumber\\
&+ 	L_0 U(dt)\rho(t)U^{\dagger}(dt)L^{\dagger}_0,
\end{align}  
where $L_0 = \sqrt{\mathcal{I}-\gamma dt\sum_{i}  L_i L^{\dagger}_i}$  is no-jump operator.  The above update can be interpreted as average over stochastic phase jump at each time step with probability $\gamma dt$. As shown in Fig. \ref{deco}, due to dephasing decay of OTOC in integrable regime almost overlaps with that of the non-integrable case for $T_2^* = 2/J$. 
  
 \bibliography{MQC}{}
 \bibliographystyle{unsrt}

  \end{document}